\documentclass[pre,twocolumn,showpacs,square,amssymb,amsmath,nobibnotes]{revtex4-1}
\usepackage{bm}
\usepackage{times}
\usepackage{graphicx}
\usepackage[usenames,dvipsnames,svgnames,table]{xcolor}
\usepackage{hyperref}
\usepackage{physics}
\hypersetup{colorlinks=true,  linkcolor=NavyBlue, citecolor=PineGreen,urlcolor=cyan}
\usepackage{physics}
\usepackage{comment}
\usepackage{xcolor}
\usepackage[normalem]{ulem}

\hypersetup{colorlinks=true, linkcolor=NavyBlue, citecolor=PineGreen,urlcolor=cyan}

\newcommand{\red}[1]{{\color{red}#1}}

\newcommand{\nn}{\nonumber}
\newcommand{\be}{\begin{equation}}
\newcommand{\ee}{\end{equation}}
\DeclareMathOperator{\sign}{sign}
\newcommand{\moy}[1]{\ensuremath{\Big\langle #1 \Big\rangle}}

\begin{document}

\global\long\def\so#1{\red{\sout{#1}}}
\global\long\def\l{\lambda}%
\global\long\def\ints{\mathbb{Z}}%
\global\long\def\nat{\mathbb{N}}%
\global\long\def\re{\mathbb{R}}%
\global\long\def\com{\mathbb{C}}%
\global\long\def\dff{\triangleq}%
\global\long\def\df{\coloneqq}%
\global\long\def\del{\nabla}%
\global\long\def\cross{\times}%
\global\long\def\der#1#2{\frac{d#1}{d#2}}%
\global\long\def\bra#1{\left\langle #1\right|}%
\global\long\def\ket#1{\left|#1\right\rangle }%
\global\long\def\braket#1#2{\left\langle #1|#2\right\rangle }%
\global\long\def\ketbra#1#2{\left|#1\right\rangle \left\langle #2\right|}%
\global\long\def\paulix{\begin{pmatrix}0  &  1\\
 1  &  0 
\end{pmatrix}}%
\global\long\def\pauliy{\begin{pmatrix}0  &  -i\\
 i  &  0 
\end{pmatrix}}%
\global\long\def\sinc{\mbox{sinc}}%
\global\long\def\ft{\mathcal{F}}%
\global\long\def\dg{\dagger}%
\global\long\def\bs#1{\boldsymbol{#1}}%
\global\long\def\norm#1{\left\Vert #1\right\Vert }%
\global\long\def\H{\mathcal{H}}%
\global\long\def\tens{\varotimes}%
\global\long\def\rationals{\mathbb{Q}}%
 
\global\long\def\tri{\triangle}%
\global\long\def\lap{\triangle}%
\global\long\def\e{\varepsilon}%
\global\long\def\broket#1#2#3{\bra{#1}#2\ket{#3}}%
\global\long\def\dv{\del\cdot}%
\global\long\def\eps{\gamma}%
\global\long\def\rot{\vec{\del}\cross}%
\global\long\def\pd#1#2{\frac{\partial#1}{\partial#2}}%
\global\long\def\L{\mathcal{L}}%
\global\long\def\inf{\infty}%
\global\long\def\d{\delta}%
\global\long\def\D{\Delta}%
\global\long\def\r{\rho}%
\global\long\def\hb{\hbar}%
\global\long\def\s{\sigma}%
\global\long\def\t{\tau}%
\global\long\def\O{\Omega}%
\global\long\def\a{\alpha}%
\global\long\def\b{\beta}%
\global\long\def\th{\theta}%
\global\long\def\l{\lambda}%

\global\long\def\Z{\mathcal{Z}}%
\global\long\def\z{\zeta}%
\global\long\def\ord#1{\mathcal{O}\left(#1\right)}%
\global\long\def\ua{\uparrow}%
\global\long\def\da{\downarrow}%
 
\global\long\def\co#1{\left[#1\right)}%
\global\long\def\oc#1{\left(#1\right]}%
\global\long\def\tr{\mbox{tr}}%
\global\long\def\o{\omega}%
\global\long\def\nab{\del}%
\global\long\def\p{\psi}%
\global\long\def\pro{\propto}%
\global\long\def\vf{\varphi}%
\global\long\def\f{\phi}%
\global\long\def\mark#1#2{\underset{#2}{\underbrace{#1}}}%
\global\long\def\markup#1#2{\overset{#2}{\overbrace{#1}}}%
\global\long\def\ra{\rightarrow}%
\global\long\def\cd{\cdot}%
\global\long\def\v#1{\vec{#1}}%
\global\long\def\fd#1#2{\frac{\d#1}{\d#2}}%
\global\long\def\P{\Psi}%
\global\long\def\dem{\overset{\mbox{!}}{=}}%
\global\long\def\Lam{\Lambda}%
 
\global\long\def\m{\mu}%
\global\long\def\n{\nu}%

\global\long\def\ul#1{\underline{#1}}%
\global\long\def\at#1#2{\biggl|_{#1}^{#2}}%
\global\long\def\lra{\leftrightarrow}%
\global\long\def\var{\mbox{var}}%
\global\long\def\E{\mathcal{E}}%
\global\long\def\Op#1#2#3#4#5{#1_{#4#5}^{#2#3}}%
\global\long\def\up#1#2{\overset{#2}{#1}}%
\global\long\def\down#1#2{\underset{#2}{#1}}%
\global\long\def\lb{\biggl[}%
\global\long\def\rb{\biggl]}%
\global\long\def\RG{\mathfrak{R}_{b}}%
\global\long\def\g{\gamma}%
\global\long\def\Ra{\Rightarrow}%
\global\long\def\x{\xi}%
\global\long\def\c{\chi}%
\global\long\def\res{\mbox{Res}}%
\global\long\def\dif{\mathbf{d}}%
\global\long\def\dd{\mathbf{d}}%
\global\long\def\grad{\vec{\del}}%

\global\long\def\mat#1#2#3#4{\left(\begin{array}{cc}
 #1  &  #2\\
 #3  &  #4 
\end{array}\right)}%
\global\long\def\col#1#2{\left(\begin{array}{c}
 #1\\
 #2 
\end{array}\right)}%
\global\long\def\sl#1{\cancel{#1}}%
\global\long\def\row#1#2{\left(\begin{array}{cc}
 #1  &  ,#2\end{array}\right)}%
\global\long\def\roww#1#2#3{\left(\begin{array}{ccc}
 #1  &  ,#2  &  ,#3\end{array}\right)}%
\global\long\def\rowww#1#2#3#4{\left(\begin{array}{cccc}
 #1  &  ,#2  &  ,#3  &  ,#4\end{array}\right)}%
\global\long\def\matt#1#2#3#4#5#6#7#8#9{\left(\begin{array}{ccc}
 #1  &  #2  &  #3\\
 #4  &  #5  &  #6\\
 #7  &  #8  &  #9 
\end{array}\right)}%
\global\long\def\su{\uparrow}%
\global\long\def\sd{\downarrow}%
\global\long\def\coll#1#2#3{\left(\begin{array}{c}
 #1\\
 #2\\
 #3 
\end{array}\right)}%
\global\long\def\h#1{\hat{#1}}%
\global\long\def\colll#1#2#3#4{\left(\begin{array}{c}
 #1\\
 #2\\
 #3\\
 #4 
\end{array}\right)}%
\global\long\def\check{\checked}%
\global\long\def\v#1{\vec{#1}}%
\global\long\def\S{\Sigma}%
\global\long\def\F{\Phi}%
\global\long\def\M{\mathcal{M}}%
\global\long\def\G{\Gamma}%
\global\long\def\im{\mbox{Im}}%
\global\long\def\til#1{\tilde{#1}}%
\global\long\def\kb{k_{B}}%
\global\long\def\k{\kappa}%
\global\long\def\ph{\phi}%
\global\long\def\el{\ell}%
\global\long\def\en{\mathcal{N}}%
\global\long\def\asy{\cong}%
\global\long\def\sbl{\biggl[}%
\global\long\def\sbr{\biggl]}%
\global\long\def\cbl{\biggl\{}%
\global\long\def\cbr{\biggl\}}%
\global\long\def\hg#1#2{\mbox{ }_{#1}F_{#2}}%
\global\long\def\J{\mathcal{J}}%
\global\long\def\diag#1{\mbox{diag}\left[#1\right]}%
\global\long\def\sign#1{\mbox{sgn}\left[#1\right]}%
\global\long\def\T{\th}%
\global\long\def\rp{\reals^{+}}%

\title{Spectral Analysis of Current Fluctuations in Periodically Driven Stochastic Systems}
\author{Bertrand Lacroix-A-Chez-Toine}
\affiliation{Department of Physics of Complex Systems, Weizmann Institute of Science,
Rehovot 7610001, Israel}
\author{Oren Raz}
\affiliation{Department of Physics of Complex Systems, Weizmann Institute of Science,
Rehovot 7610001, Israel}

\begin{abstract}
Current fluctuations play an important role in non-equilibrium statistical mechanics, and are a key object of interest in both theoretical studies and in practical applications. So far, most of the studies were devoted to the fluctuations in the time-averaged current -- the zero frequency Fourier component of the time dependent current. However, in many practical applications the fluctuations at other frequencies are of equal importance. Here we study the full frequency dependence of current statistics in periodically driven stochastic systems. First, we show a general method to calculate the current statistics, valid even when the current's frequency is incommensurate with the driving frequency, breaking the time periodicity of the system. Somewhat surprisingly, we find that the cumulant generating function (CGF), that encodes all the statistics of the current, is composed of a continuous background at any frequency accompanied by either positive or negative peaks at current's frequencies commensurate with the driving frequency. We show that cumulants of increasing orders display peaks at an increasing number of locations but with decreasing amplitudes that depend on the commensurate ratio of frequencies. All these peaks are then transcribed in the behaviour of the CGF. As the measurement time increases, these peaks become sharper but keep the same amplitude and eventually lead to discontinuities of the CGF at all the frequencies that are commensurate with the driving frequency in the limit of infinitely long measurement. 
We demonstrate our formalism and its consequences on three types of models: an underdamped Brownian particle in a periodically driven harmonic potential; a periodically driven run-and-tumble particle; and a two-state system. 
\end{abstract}

\maketitle

\section{Introduction}

Equilibrium statistical mechanics is a mature framework that provides a calculation scheme for the ensemble averages and the fluctuations of many quantities of interest, with only mild assumptions on the system, e.g. thermodynamic equilibrium and short range interactions. 
In practice, however, equilibrium systems are rather the exception than the norm. Despite many efforts, no equivalent general framework is known for systems which are far from thermal equilibrium. Characterising the mean value and fluctuations for most quantities of interest remains a challenging task, even in the relatively simple cases of a system in a non-equilibrium steady state \cite{zia2007probability,ruelle2003extending,zia2006possible} or a periodically driven system \cite{brandner2015thermodynamics,raz2016mimicking,busiello2018similarities}.

 Important progress has been achieved in the 50's for systems which are close to thermal equilibrium, in the form of the fluctuation-dissipation relations \cite{kubo1966fluctuation}. These relate the equilibrium fluctuations of a system to the rate of dissipation when the system is weakly driven away from equilibrium. These fluctuation-dissipation relations where generalized in the 90's to the Gallavotti-Cohen fluctuation relations, which constitute rare exact results holding arbitrarily far from equilibrium. They imply a fundamental symmetry on the distributions of the entropy production \cite{gallavotti1995dynamical,kurchan1998fluctuation} and of the currents \cite{Gaspard2007fluctuation} in the system and naturally extend the fluctuation dissipation relations. They hold for non-equilibrium steady state systems \cite{gallavotti1995dynamical,lebowitz1999gallavotti,andrieux2007fluctuation} as well as periodically driven systems \cite{shargel2009fluctuation}, although they are not expected to hold for currents in the latter case \cite{Harris_2007}. 
 
The thermodynamic uncertainty relations, bounding the current fluctuations in the system using the entropy production \cite{barato2015thermodynamic,gingrich2016dissipation,horowitz2019thermodynamic} constitute another important progress that was recently achieved and holds arbitrarily far from equilibrium. They imply that decreasing the current fluctuations in the steady state comes at a cost in terms of the entropy production. These relations were extended recently to periodically driven systems in \cite{proesmans2019hysteretic,barato2018bounds,koyuk2019operationally,PhysRevLett.125.260604,koyuk2018generalization}. Both the Gallavotti-Cohen and the thermodynamic uncertainty principle hold for arbitrary systems, but in practice they are of most interest in the case of small systems where large fluctuations occur with a non-vanishing probability.

In this study we consider the fluctuations of the current's Fourier modes in stochastic non-equilibrium systems that are subject to time-periodic driving. The fluctuations of the zero frequency current (commonly referred to as the ``DC current") are tightly related to both the Gallavotti-Cohen and the thermodynamic uncertainty relations described above, and thus were studied intensively for systems which are subject to a time-independent thermodynamic forcing that breaks detailed balance, driving them into a non-equilibrium steady state \cite{nyawo2016large,nemoto2011variational,nemoto2011thermodynamic,proesmans2019large,derrida2019large,nyawo2017minimal,fischer2018large}. However, only few works have been dedicated to other Fourier modes of the currents. To the best of our knowledge,  only the zero-frequency current fluctuations \cite{barato2018current,bertini2018level} and the fluctuations of the current's Fourier mode with the same frequency as that of the drive \cite{Raphael2020periodically} have been investigated.

The motivation to study all of the current's Fourier modes is natural from a theoretical perspective, but also from a practical one, as often these fluctuations have important implications. An interesting example is the Paul trap, commonly used to trap ions \cite{blatt2008entangled}. Ions in this trap display heating processes commonly orders of magnitude higher than expected \cite{brownnutt2015ion} which are yet poorly understood. In this system ions are trapped by imposing an alternating current at high frequency giving also rise to fluctuations at all other frequencies. Although the driving frequency is far from the frequency that can heat the ions, the generated current fluctuation at the ion's frequency might be an important component in the heating process.

In this manuscript we study the frequency dependence of current statistics when a stochastic Markovian system is periodically driven. Our main object of interest is the \emph{Cumulant generating function} (CGF) of the current Fourier modes, which encodes the full information on the current statistics. First, we use the Oseledets theorem to extend the framework of the current cumulant generating function from the case where the driving frequency is commensurate with the Fourier mode of the current, to account for non-commensurate frequency ratios as well. Surprisingly, we find that this function has a structure similar to the famous Thomae's function \cite{thomaes}, namely it is composed of a continuous ``background" for frequencies which are incommensurate with respect to the periodic driving, accompanied by (positive or negative) discontinuous ``spikes" at commensurate frequencies. The integer denominator of the commensurate ratio between the frequencies at which the spikes appear can be directly related to the order of the cumulant that has a discontinuous behaviour at this frequency. Moreover, we explain how our results are modified for a finite time experiment, where the discontinuous spikes are smoothed into continuous peaks that gets sharper as the duration of the measurement increases. This general behavior is demonstrated on a Brownian particle in a periodically driven harmonic potential and a periodically driven trapped run-and-tumble particle, where analytical results for the first two cumulants can be obtained, and in a two-state system where the cumulant generating function can be numerically evaluate.

The setup studied in this manuscript is somewhat similar to the \emph{stochastic resonance} setup, where a bi-stable system coupled to a noise source is weakly perturbed by time-periodic forcing (See \cite{gammaitoni1998stochastic} for a review on stochastic resonance). However, the main object of interest in stochastic resonance is the response of the system to the driving as a function of the noise, where the response of the system is usually the transitions between the two meta-stable states at the driving frequency. In other words, the interest in stochastic resonance is in the driving frequency Fourier mode of the current between the two meta-stable state. In contrast, we consider a quite generic periodic driving, which is not necessarily weak and does not necessarily have two meta-stable states. We also consider all the Fourier modes of the current statistics, and not only the average current as in stochastic resonance. Furthermore, we are interested in the frequency dependence of the fluctuations.

The manuscript is organised as follows: In section \ref{setup} we detail the general setting that we consider and give the main technical background needed to obtain our results. In section \ref{main_results} we provide a short summary of our main results and their implications. In section \ref{der_cum} we derive our main results by obtaining general expressions of the cumulants. In section \ref{examples} we show how to apply our results to simple exactly solvable systems. Finally in section \ref{conclu} we conclude and give some insights on new unexplored directions. Some technical details of the computations and complicated expressions are relegated to the Appendices.

\section{Setup -- Periodically driven Markovian systems}\label{setup}

\subsection{Markov propagator}

In this manuscript we consider the statistics of the current's Fourier modes in a periodically driven stochastic system.  To this end, we assume that the evolution of the probability distribution associated with the system is Markovian, as detailed in what follows. The micro-state of the system, for example the position of a particle in space, is denoted by the vector ${\bf x}$. We assume that the space of all ${\bf x}$ is simply connected \footnote{This assumption implies some consequences on our results. Although a generalization to non-simply connected domains is possible, we do not consider these cases here}. The probability distribution of all the micro-states at time $t$ is uniquely characterised by the initial state of the system, which for simplicity we assume to be a specific micro-state ${\bf x_0}$, as well as by the propagator, $G({\bf x},t|{\bf x}_0,t_0)$, which is the  transition probability density to be in a final state ${\bf x}$ at time $t$, provided that the system was at ${\bf x}_0$ at time $t_0$. It satisfies the evolution equation,
\begin{align}
&\partial_{t} G={\cal L}({\bf x}, \nabla_{\bf x},t)G\,,\label{FP_0}\\
&{\cal L}({\bf x},\nabla_{\bf x},t)=-\sum_{\beta}\partial_{x_\beta} F_\beta({\bf x},t)+\sum_{\beta,\gamma}\partial_{x_\beta,x_\gamma} D_{\beta\gamma}({\bf x},t) \;,\nn
\end{align}
where here and in the following we use Greek letters for the indices of the micro-state space and with the initial condition
\begin{equation}
    \,\,G({\bf x},t_0|{\bf x}_0,t_0)=\delta^d({\bf x}-{\bf x}_0)\,,
\end{equation}
where $\delta^{d}({\bf x})=\prod_{\beta=1}^d \delta(x_\beta)$ is the $d$-dimensional Dirac delta function. By virtue of probability conservation, the propagator is normalised to unity,
\be
\forall t\geq t_0\;,\;\;\int d{\bf x}\, G({\bf x},t|{\bf x}_0,t_0)=1\;.
\ee

Throughout the manuscript we assume that the force $F_\beta({\bf x},t)$ and diffusion $D_{\beta \gamma}({\bf x},t)$ are both explicitly time dependent, non-singular and  time-periodic with cycle time $T_d$. We denote the corresponding fundamental angular frequency by $\omega_d=(2\pi)/T_d$, and we refer to it as the \emph{driving frequency}, as this time dependence is what drives the system out of equilibrium. We are interested in the cases where the system has a unique time-periodic stationary state in the limit $t\to \infty$, and that it does not depend on the initial position ${\bf x}_0$.

Equation \eqref{FP_0} can be formally solved by
\be
G({\bf x},t|{\bf x}_0,t_0)={\cal T}e^{\int_{t_0}^{t}d\tau {\cal L}({\bf x},\nabla_{\bf x},\tau)}\delta^d({\bf x}-{\bf x}_0)\;,
\ee
where ${\cal T}$ is the time-ordered product. Exploiting the  discrete time translation symmetry of the system, ${\cal L}({\bf x},\nabla_{\bf x},t+T_d)={\cal L}({\bf x},\nabla_{\bf x},t)$ for any time $t$, we expand the propagator using Floquet theory as \cite{caceres2006theory,kim2010rate}
\be
G({\bf x},t|{\bf x}_0,t_0)=\sum_{k=0}^{\infty} e^{-\lambda_k (t-t_0)}f_k({\bf x},t)g_k({\bf x}_0,t_0)\;,\label{prop_floquet}
\ee
where the functions $f_k$ and $g_k$ are 
the eigenvectors $U_{T_d} f_k=e^{-\lambda_k T_d}f_k$ and $U_{T_d}^{\dagger} g_k=e^{\lambda_k T_d}g_k$ of the operators
\begin{align}
    &U_{T_d}(t)={\cal T}e^{\int_{t}^{t+T_d}d\tau {\cal L}({\bf x},\nabla_{\bf x},\tau)}\;,\\
    &U_{T_d}^{\dagger}(t_0)={\cal T}e^{-\int_{t_0}^{t_0+T_d}d\tau {\cal L}^{\dagger}({\bf x}_0,\nabla_{{\bf x}_0},\tau)}\;,
\end{align}
where
\be
{\cal L}^{\dagger}({\bf x},\nabla_{\bf x},t_0)=\sum_{\beta}F_\beta({\bf x},t_0)\partial_{x_\beta} +\sum_{\beta,\gamma} D_{\beta\gamma}({\bf x},t_0)\partial_{x_\beta,x_\gamma}
\ee
is the adjoint of the evolution operator ${\cal L}({\bf x},\nabla_{\bf x},t)$.
 The Floquet eigenvalues $\lambda_k$'s are ordered in the following way, 
\be
\Re(\lambda_0)=0< \Re(\lambda_1)\leq \Re(\lambda_2)\leq \cdots
\ee
and they all have positive real part. Using our assumptions on the periodic solution being unique, we obtain that the eigenvalue $\lambda_0=0$ is non-degenerate. 
On the other hand, as $(t-t_0)\to \infty$, we obtain that 
\be
G({\bf x},t|{\bf x}_0,t_0)\xrightarrow[t\rightarrow\infty]{} f_0({\bf x},t)g_0({\bf x}_0,t_0)\;.
\ee
As this stationary state is independent of ${\bf x}_0$, we must have that $g_0({\bf x}_0,t_0)$ is independent of both ${\bf x}_0$ and $t_0$. It is convenient to choose $g_0({\bf x}_0,t_0)=1$, so that the associated eigenfunction $f_0({\bf x},t)$ is positive and normalised to unity. Note that we assume this steady state to be non-singular, i.e. $f_0({\bf x},t)>0$ at least in a finite domain.   More generally, exploiting the identities
\begin{align}
&G({\bf x},t|{\bf x}_0,t)=\delta^d({\bf x}-{\bf x}_0)\\
&\int d{\bf y}\,G({\bf x},t|{\bf y},\tau)G({\bf y},\tau|{\bf x}_0,t_0)=G({\bf x},t|{\bf x}_0,t_0)
\end{align}
one can show that the eigenfunctions $f_k({\bf x},t)$ and $g_k({\bf x}_0,t)$ are bi-orthogonal and satisfy a completeness relation 
\begin{align}
    &\sum_k f_k({\bf x},t)g_k({\bf x}_0,t)=\delta^d({\bf x}-{\bf x}_0)\;,\\
    &\int d{\bf x}\, f_k({\bf x},t)g_l({\bf x},t)=\delta_{k,l}\;,   \label{bi_ortho} 
\end{align}
where we remind that $\delta^{d}({\bf x})=\prod_{\beta=1}^d \delta(x_\beta)$ is the $d$-dimensional Dirac delta function while $\delta_{k,l}=1$ if $k=l$ and zero otherwise is the Kronecker delta.

\subsection{Alternating charge}

So far we discussed the probability propagator in periodically driven systems. However, in this manuscript we are interested in the fluctuations of the oscillating current at an arbitrary frequency $\omega_c$. In practice, one needs to consider the associated time-integrated quantity which, in analogy to the DC case, we denote by the {\it fluctuating charge}. For a given realisation, the empirical $\omega_c$-alternating charge is defined as
\begin{align}
&Q_{\omega_c;\alpha}(t)=\int_0^t d\tau\,\partial_{\tau} x_\alpha(\tau)\,\cos(\omega_c \tau)\label{def_AC}\\
&=\left[x_\alpha(\tau)\cos(\omega_c \tau)\right]_{\tau=0}^{\tau=t}+\omega_c\int_0^t d\tau\,x_\alpha(\tau)\,\sin(\omega_c \tau)\;,\nn
\end{align}
where $x_\alpha(\tau)$ is the $\alpha^{\rm th}$ component of ${\bf x}(\tau)$ \footnote{Note that a similar quantity can be defined with a $\sin$ function instead of the $\cos$.}. The random variable $Q_{\omega_c;\alpha}(t)$ is a functional of the whole path realisation ${\bf x}(\tau)$ for all $\tau\in[0,t]$. However,  $Q_{\omega_c;\alpha}(t)$ can be calculated for any specific realisation of the system using the above integral.

For $\omega_c=0$, the empirical alternating charge coincides with the direct empirical charge. As we consider a simply-connected space, this direct charge is path independent, given the values of the initial $x_{\alpha}(0)$ and final $x_{\alpha}(t)$ positions, and is equal to $Q_{\omega_c=0;\alpha}(t)=x_{\alpha}(t)-x_{\alpha}(0)$. The probability distribution function (PDF)  $P_{\omega_c=0,\alpha}(Q;t)$ 
 of this direct charge $Q_{0,\alpha}(t)$ is given, in the large time limit, by 
\be
P_{\omega_c=0,\alpha}(Q;t)\to \int d{\bf x}\, f_0({\bf x},t)\delta(x_{\alpha}-x_{0,\alpha}-Q)\;.
\ee

For a generic value of the frequency $\omega_c$, the empirical alternating charge is path dependent, and therefore  computing the probabilty distribution function $P_{\omega_c,\alpha}(Q;t)$ of the alternating charge is a non-trivial task. It turns out to be simpler in this case to consider instead its moment generating function (MGF) 
\begin{align}
\moy{e^{\mu Q_{\omega_c;\alpha}(t)}}_{{\bf x}_0}&\equiv\int_{-\infty}^{\infty} dQ\, e^{\mu Q} P_{\omega_c,\alpha}(Q;t),
\end{align}
where the average $\langle\cdots\rangle_{{\bf x}_0}$ is taken over all the stochastic trajectories starting from the same initial state ${ \bf x}(0)={\bf x}_0$. This generating function can also be written as
\begin{align}
\moy{e^{\mu Q_{\omega_c;\alpha}(t)}}_{{\bf x}_0}&=\int d{\bf x} \,G_{\mu}({\bf x},t|{\bf x}_0)\,,
\end{align}
where the function $G_{\mu}({\bf x},t|{\bf x}_0)$ evolves in time according to the so-called "tilted" operator \cite{chetrite2013nonequilibrium}
\begin{align}
&\partial_{t} G_\mu={\cal L}_{\mu}({\bf x},\nabla_{\bf x},t)G_\mu\,,\\
&G_\mu({\bf x},0|{\bf x}_0)=\delta^d({\bf x}-{\bf x}_0)\,.
\end{align}
The tilted operator can be written as
\begin{align}
&{\cal L}_{\mu}({\bf x},\nabla_{\bf x},t)=\sum_{k=0}^2 \left[\mu\cos(\omega_c t)\right]^k {\cal L}^k({\bf x},\nabla_{\bf x},t)\;,
\end{align}
where
\begin{align}
&{\cal L}^0={\cal L}({\bf x},\nabla_{\bf x},t)\;,\\
&{\cal L}^1=F_\alpha({\bf x},t)-\sum_{\beta}\partial_{x_{\beta}}\left(D_{\alpha\beta}({\bf x},t)+D_{\beta\alpha}({\bf x},t)\right)\;,\nn\\
&{\cal L}^2=D_{\alpha\alpha}({\bf x},t)\;.\nn
\end{align}
Similarly to the case of the propagator, one can formally express the solution as
\be
G_{\mu}({\bf x},t|{\bf x}_0)={\cal T}e^{\int_0^t d\tau {\cal L}_{\mu}({\bf x},\nabla_{\bf x},t)}\delta^d({\bf x}-{\bf x}_0)\;.
\ee
We note, however, that in contrast with ${\cal L}({\bf x},\nabla_{\bf x},t)$ which is periodic by construction, ${\cal L}_{\mu}({\bf x},\nabla_{\bf x},t)$ is not necessarily time periodic, as it has both periodic components varying with frequency $\omega_c$ and frequency $\omega_d$. Only in the case where the ratio between these frequencies is commensurate, namely when it is given by a rational number, does ${\cal L}_{\mu}({\bf x},\nabla_{\bf x},t)$ become periodic. In what follows, we first address the commensurate case, and then the more challenging case of non-commensurate frequencies.

\subsection{Commensurate frequencies}

Let us first consider the case where the frequencies of the drive $\omega_d$ and of the charge $\omega_c$ are commensurate. In this case there exist two natural numbers, $n,m\in\mathbb{N}$, such that
\be
T=n T_d=m T_c<\infty\;.
\ee
In this case the tilted operator is periodic with a finite period $T$, namely ${\cal L}_{\mu}({\bf x},\nabla_{\bf x},t+T)= {\cal L}_{\mu}({\bf x},\nabla_{\bf x},t)$. One may then use the Floquet theory by defining the operators
\begin{align}
    &U_{T;\mu}(t)={\cal T}e^{\int_{t}^{t+T}d\tau {\cal L}_{\mu}({\bf x},\nabla_{\bf x},\tau)}\;,\\
    &U_{T;\mu}^{\dagger}(t_0)={\cal T}e^{-\int_{t_0}^{t_0+T}d\tau {\cal L}_{\mu}^{\dagger}({\bf x}_0,\nabla_{{\bf x}_0},\tau)}\;.
\end{align}
Introducing the common ordered set of Floquet eigenvalues of these operators $\{\lambda_{k}(\mu)\,,\,\,\Re[\lambda_0(\mu)]<\Re[\lambda_{1}(\mu)]\leq \cdots\}$ and their associated respective eigenfunctions $f_k^{\mu}({\bf x},t)$ and $g_k^{\mu}({\bf x}_0,t)$, we may express the MGF as
\be
\moy{e^{\mu Q_{\omega_c;\alpha}(t)}}=\sum_{k}e^{-\lambda_k(\mu) t}a_k(t;{\bf x}_0)\;,\label{MGF_lamb}
\ee
 where the functions $a_k(t;{\bf x}_0)=\int d{\bf x}f_k^{\mu}({\bf x},t)g_k^{\mu}({\bf x}_0,0)$ are periodic in $t$ with period $T$. In contrast to the propagator $G({\bf x},t|{\bf x}_0)$, the tilted propagator $G_{\mu}({\bf x},t|{\bf x}_0)$ does not propagate a probability distribution, and its largest eigenvalue $\lambda_0(\mu)$ is non zero in general, apart for $\mu=0$ where ${\cal L}_{\mu=0}({\bf x},\nabla_{\bf x},t)={\cal L}({\bf x},\nabla_{\bf x},t)$. Using the Floquet theory, the {\it cumulant generating function} (CGF) in the large time limit reads
\begin{align}
\chi_{\mu}(\omega_c)&\equiv\lim_{t\to \infty}\frac{1}{t}\ln\moy{e^{\mu Q_{\omega_c;\alpha}(t)}}_{\bf x_0}=-\lambda_0(\mu)\label{def_CGF}\\
&=\sum_{p=1}^{\infty}{\cal Q}_{p}^{\alpha}(\omega_c)\, \mu^p\;,\nn
\end{align}
where ${\cal Q}_{p}^{\alpha}(\omega_c)$ corresponds to the large time limit of the rescaled cumulant of the alternating charge 
\be
{\cal Q}_{p}^{\alpha}(\omega_c)\equiv\lim_{t\to \infty}\frac{\moy{Q_{\omega_c;\alpha}^p(t)}_{\rm co}}{t}=-\left.\partial_\mu^p\lambda_0(\mu)\right|_{\mu=0}\;.
\ee
Here, the subscript $_{\rm co}$ refers to a connected correlation function, and $p$ is the order of the cumulant. The limit is well-defined as the cumulants scale at most linearly with time.
 The specific form of the CGF obtained here is consistent with the probability distribution function (PDF) $P_{\omega_c}(Q;t)$ of the alternating charge $Q_{\omega_c;\alpha}(t)$ taking a large deviation scaling form in the large time limit. By definition, the MGF 
\be
\moy{e^{\mu Q_{\omega_c;\alpha}(t)}}=\hat{ P}_{\omega_c}(-i \mu;t)\;,
\ee
where $\hat{ P}_{\omega_c}(k;t)$ is the Fourier transform of the PDF $P_{\omega_c}(Q;t)$ at the specific value $k=-i \mu$. Taking the inverse Fourier transform, we obtain that the {\it atypical fluctuations} of the alternating charge are described by the large deviation form,
\be
P_{\omega_c}(Q;t)=\int_{-i \infty}^{i\infty} \frac{d\mu}{2i \pi}e^{-\mu Q}\moy{e^{\mu Q_{\omega_c;\alpha}(t)}} \asymp e^{-t\, \Phi_{\omega_c}\left(\frac{Q}{t}\right)}\;,\label{large_dev}
\ee
where we have used the notation (common in large deviation theory)
$$A(x,t)\asymp e^{-t B\left(\frac{x}{t}\right)}\Leftrightarrow-\lim_{t\to \infty }\frac{1}{t}\ln A(at,t)=B\left(a\right)\;.$$  
The large deviation function $\Phi_{\omega_c}(J)$ is given by the Legendre transform of the CGF
\be
\Phi_{\omega_c}(J)=\max_{\mu}\left[\mu\, J-\chi_{\omega_c}(\mu)\right]\;,
\ee
which is simply obtained via a saddle point approximation for large $t$ with $J=Q/t<\infty$ as $t\to \infty$. Furthermore, as all the non-zero cumulants have the same scaling $O(t)$, we also expect the central limit theorem to hold for the long-time {\it typical fluctuations} of the alternating charge. The corresponding distribution is Gaussian with fluctuations of order $\sqrt{t}$ around the average value
\be
P_{\omega_c,\alpha}(Q;t)\approx \frac{\displaystyle\exp\left(-\frac{(Q-\moy{Q_{\omega_c;\alpha}(t)})^2}{2{\rm Var}(Q_{\omega_c;\alpha}(t))}\right)}{\sqrt{2\pi{\rm Var}(Q_{\omega_c;\alpha}(t))}}\;.
\ee

\subsection{Incommensurate frequencies}

We now consider the situation where the frequency $\omega_d$ of the drive and $\omega_c$ of the current are incommensurate.
As the tilted operator ${\cal L}({\bf x},\nabla_{\bf x},t)$ is {\it not} periodic, we cannot use a Floquet expansion to describe the tilted propagator $G_{\mu}({\bf x},t|{\bf x}_0)$ and it is not obvious how to compute the limit in equation \eqref{def_CGF}.
However, one can use Oseledets' multiplicative ergodic theorem \cite{oseledets1968multiplicative} to show, nevertheless, that this limit is well defined for any frequency $\omega_c$. This implies in particular that all the cumulants scale at most linearly in time in this more generic case. 

A naive way to compute the CGF in this case is to consider a series of frequencies $\omega_c(n)$, commensurate with $\omega_d$ for any finite $n$ but that converges to the desired incommensurate value of $\omega_c$ as $n\to \infty$. The CGF can be calculated for any $n$ using the technique described above for commensurate ratios, and in the limit $n\to\infty$ one can hope to estimate $\chi_\mu(\omega_c)$. However, this naive approach only works if the CGF $\chi_{\mu}(\omega_c)$ is a continuous function of $\omega_c$ (at a fixed value of $\mu$), which we prove below quite surprisingly not to be the case. Therefore, at incommensurate frequency ratios the CGF can (to the best our knowledge) only be evaluated by taking the $t\to\infty$ limit in equation \eqref{def_CGF}.

\section{Main results}\label{main_results}

Let us summarise our main findings and their consequences. Our first main result is that for a periodically driven system with driving frequency $\omega_d$, the cumulant generating function (CGF)  of the time-integrated alternating current $Q_{\omega_c;\alpha}(t)$, namely $\chi_\mu(\omega_c)$ defined in in Eq. \eqref{def_CGF}, is \emph{not} a continuous function of the current's frequency $\omega_c$ in the large time limit.
The CGF varies smoothly for any current frequency $\omega_c$ incommensurate with the frequency $\omega_d$ of the drive but displays additional discontinuous spikes at commensurate frequencies. This feature can be explained by considering the behaviour of the rescaled cumulants ${\cal Q}_{p}^{\alpha}(\omega_c)$ of order $p\geq 1$ as a function of the current frequency $\omega_c$. The cumulants of even order, namely with $p=2r$ and $r\in \mathbb{N}^*$, have a generically non-zero continuous background for irrational frequency ratios $\omega_c/\omega_d\notin \mathbb{Q}$ and they display discontinuous spikes at frequencies
\be
\omega_c=\frac{n}{2z}\omega_d\;,\;\;n\in\mathbb{N}^*\;,\;\;1\leq z\in\mathbb{N}\leq r\;.\label{Eq.Even_p_Spikes}
\ee
The cumulants of odd order, with $p=2r+1$ and $r\in \mathbb{N}$, have zero continuous background for irrational ratios $\omega_c/\omega_d\notin \mathbb{Q}$, but they do display discontinuous spikes at frequencies
\be
\omega_c=\frac{n}{2z+1}\omega_d\;,\;\;n\in\mathbb{N}^*\;,\;\;0\leq z\in\mathbb{N}\leq r\;.\label{Eq.Odd_p_Spikes}
\ee
Note that the number of spikes increases with the order $p$ of the cumulant. However, the height of these spikes decreases with the order $p$, and goes to zero at the $p\to \infty$ limit. 
A consequence of this result is that for any fixed $\omega_c$ which is incommensurate with $\omega_d$, the CGF is symmetric $\chi_{\omega_c}(\mu)=\chi_{\omega_c}(-\mu)$, implying an analogous symmetry of the PDF $P_{\omega_c}(Q,t)=P_{\omega_c}(-Q,t)$ in the large time limit. -

The discontiuous behavior of the cumulant generating function is a result of the long time limit: one clearly does not expect any discontinuity of the cumulant generating function at finite time. We find that at large, but finite time $t$, the behavior of the cumulant generating function is qualitatively the same, but as expected the spikes in the spectrum of the cumulants are smeared over a typical frequency scale of order $\sim t^{-1}$.

These results are illustrated by considering in details two exactly solvable examples and one example that can be calculated numerically. First, we consider an underdamped Brownian particle trapped in a periodically varying harmonic potential. Then, we obtain explicit results for a harmonically confined "active" particle, i.e. subject to telegraphic noise. Lastly, we consider a time-periodic two-state system, where the cumulant generation function can be numerically evaluated.

\section{Derivation of the cumulants}\label{der_cum}
We consider in this section the cumulants of the alternating charge $Q_{\omega_c;\alpha}(t)$ in the large time limit for a gaped system, i.e. $\Re(\lambda_1)>\lambda_0=0$ where $\lambda_0,\lambda_1$ are the two lowest Floquet eigenvalues associated to the "bare" evolution operator ${\cal L}({\bf x},\nabla_{\bf x},t)$, i.e. in the absence of tilting $\mu=0$ (in this case the operator is periodic by definition and therefore the Floquet thery can be applied). As seen in section \ref{setup}, we expect that the rescaled cumulants 
\be
{\cal Q}_{p}^{\alpha}(\omega_c)=\lim_{t\to \infty}\frac{\moy{Q_{\omega_c;\alpha}^p(t)}_{\rm co}}{t}\;,
\ee
are of order $O(1)$ at most (we recall that a function of time $A(t)$ is defined to be of order $O(t^{\alpha})$ if $0<\lim_{t\to \infty}t^{-\alpha} A(t)<\infty$). In the following, we derive systematically for any value of $p$ its expression as a function of the parameters $\omega_c,\omega_d$ and the Floquet spectrum $\{\lambda_k, k\in \mathbb{N}\}$. Before deriving this expression, we first use Eq. \eqref{def_AC} to rewrite 
\begin{align}
Q_{\omega_c;\alpha}(t)&=\left[x_{\alpha}(\tau)\,\cos(\omega_c \tau)\right]_0^t+q_{\omega_c;\alpha}(t)\nn\\
q_{\omega_c;\alpha}(t)&=\omega_c\int_0^{t}d\tau \, x_{\alpha}(\tau)\,\sin(\omega_c \tau)\;.
\end{align}
For the systems that we are considering, we expect the average value and fluctuations of $x_{\alpha}(t)$ to remain of $O(1)$ as $t\to \infty$. As shown in App. \ref{app_q_fin_pos}, this simplifies the computation of the cumulants for any value of $p$ as
\be
{\cal Q}_{p}^{\alpha}(\omega_c)=\lim_{t\to \infty}\frac{\moy{q_{\omega_c;\alpha}^p(t)}_{\rm co}}{t}\;. \label{Eq:Q_p^alpha}
\ee
From this identification, we already obtain that the system does not have a direct charge ${\cal Q}_{p}^{\alpha}(\omega_c=0)=0$ for any $p\geq 0$. This is indeed expected, as  non-zero cumulants of the charge would imply some fluctuations of the steady state current, which are incompatible with our assumption of simply-connected micro-state space. We note, however, that in a compact domain and a non-trivial topology, e.g. a particle on a ring, the empirical alternating and direct charges are not defined as per the second line of equation \eqref{def_AC}. Indeed, in that case, the fluctuation of the direct charge might be non-zero for such systems \cite{busiello2018similarities}.

The cumulant of order $p$ of $q_{\omega_c;\alpha}(t)$ reads
\begin{align}
&\moy{q_{\omega_c;\alpha}^p(t)}_{\rm co}=\label{cum_q_1}\moy{\prod_{j=1}^p \int_0^t dt_j\, x_\alpha(t_j)\,\omega_c\,\sin(\omega_c t_j)}_{\rm co}\\
&=p!\, \omega_c^p \int_0^{t}dt_p\cdots \int_0^{t_2}dt_1 \moy{\prod_{j=1}^p x_\alpha(t_j)}_{\rm co}\prod_{j=1}^p\sin(\omega_c t_j)\,,\nn
\end{align}
where the times $t\geq t_p\geq t_{p-1}\geq \cdots \geq t_2\geq t_1\geq 0$ in the second line are ordered. In this expression the $p$-times connected correlation functions $\moy{\prod_{j=1}^p x_{\alpha}(t_j)}_{\rm co}$ are conveniently expressed in terms of the lower order $n\leq p$-times (disconnected) correlation functions $\moy{\prod_{j=1}^n x_{\alpha}(t_j)}$ as follows \cite{speed1983cumulants}
\begin{align}
   &\moy{\prod_{k=1}^p x_{\alpha}(t_k)}_{\rm co}=\nn\\
   &\sum_{k=1}^{p} (k-1)!(-1)^{k+1}\sum_{\pi \in P_k(p)}\prod_{B\in \pi}\prod_{i=1}^k \moy{\prod_{j\in B_i} x(t_{j})}\;,\label{con_correl}
\end{align}
where $\pi\in P_k(p)$ is an element of the groups of partitions of $\{1,\cdots,p\}$ into $k$ blocks $B_i$'s with $i=1,\cdots,k$. We denote $n_i$ the number of elements in block $i$ of the partition $\pi$, with $\sum_{i=1}^k n_i=p$, and each element within a given block is ordered, i.e $B_i(1)<\cdots< B_i(n_i)$. 

The $n$-times (disconnected) correlation functions $\moy{\prod_{j=1}^n x_{\alpha}(t_j)}$ can be computed explicitly by introducing the propagator between each of the times $t_j$'s, which results in 
\begin{align}
&\moy{\prod_{j=1}^p x_{\alpha}(t_j)}\label{dis_correl}\\
&=\int d{\bf x}_1\cdots\int d{\bf x}_p \prod_{j=1}^{p}\left[x_{j,\alpha}\,G({\bf x}_j,t_{j}|{\bf x}_{j-1},t_{j-1})\right]\nn\\
&=\sum_{l_1,\cdots,l_p=0}^{\infty}\left[\prod_{j=1}^{p}e^{-\lambda_{l_j}(t_j-t_{j-1})}\right]C_{l_1,\cdots,l_n}(t_1,\cdots,t_p)\;,\nn
\end{align}
where $t_0=0$. The third line is obtained by inserting the Floquet expansion \eqref{prop_floquet} of the propagator where $l_1,\cdots,l_p$ refer to indices in the Floquet spectrum and the function 
\begin{align}
&C_{l_1,\cdots,l_p}(t_1,\cdots,t_n)=g_{l_1}({\bf x}_0,0)\int d{\bf x}_p\, x_{p,\alpha}\,f_{l_p}({\bf x}_p,t_p)\nn\\
&\times \prod_{j=1}^{p-1} \int d{\bf x}_j\, x_{j,\alpha}\,f_{l_j}({\bf x}_j,t_j)g_{l_{j+1}}({\bf x}_j,t_j)\;,
\label{C_coeff}
\end{align}
is periodic with fundamental frequency $\omega_d$ in all its variables.
Notice that as $g_0({\bf x},t)=1$ for all ${\bf x}$ and $t$, the term $\int d{\bf x}_j\, x_{j,\alpha} f_{l_j}({\bf x}_j,t_j)g_{l_{j+1}}({\bf x}_j,t_j)$ connecting $l_j$ with $l_{j+1}$ effectively becomes independent of $l_{j+1}$ for $l_{j+1}=0$. Thus this term simplifies as a product of smaller order correlation function 
\begin{align}
    &C_{l_1,\cdots,l_p}(t_1,\cdots,t_n)=\label{simpl_correl}\\
    &C_{l_1,\cdots,l_j}(t_1,\cdots,t_j)\times C_{0,\cdots,l_p}(t_{j+1},\cdots,t_n)\;,\;\;l_{j+1}=0\;.\nn
\end{align}
Note also that as $\lambda_{0}=0$ in equation \eqref{con_correl} for $l_{j+1}=0$, the product
\begin{align}
    &\prod_{m=1}^{p}e^{-\lambda_{l_m}(t_m-t_{m-1})}=\\
    &\prod_{m=1}^{j}e^{-\lambda_{l_m}(t_m-t_{m-1})}\prod_{n=j+2}^{p}e^{-\lambda_{l_n}(t_n-t_{n-1})}\;,\;\;l_{j+1}=0\;,\nn
\end{align}
effectively disconnecting the correlations before $t_{j+1}$ from the correlations after this time.


Using the above and inserting \eqref{dis_correl}, \eqref{C_coeff} into equation \eqref{con_correl}, taking the limit where $t_1\gg \lambda_1^{-1}$ and $\delta t_i=t_i-t_{i-1}=O(1)$ for $i=2,\cdots,p$, the connected correlation function simplifies to
\begin{align}
&\moy{\prod_{j=1}^p x_{\alpha}(t_j)}_{\rm co}\approx\sum_{k=1}^{p} (k-1)!(-1)^{k+1}\sum_{\pi \in P_k(p)}\prod_{B\in \pi}\label{con_correl}\\
&\sum_{l_1,\cdots,l_{p}=0}^{\infty}(1-\delta_{l_p,0})\delta_{l_1,0}\prod_{j=2}^p e^{-\mu_j(l_2,\cdots,l_p;\pi)\delta t_j}\nn \\
&\times\prod_{i=1}^k\left[\tilde C_{l_{B_i(1)},\cdots,l_{B_i(n_i)}}(t_{B_i(1)},\cdots,t_{B_i(n_i)})\right]\;,\nn
\end{align}
where, making explicit use of relation \eqref{simpl_correl}, we have introduced the functions $\tilde C_{l_1}(t_1)=\delta_{l_1,0}C_0(t_1)$ and for $p>1$
\begin{align}
&\tilde C_{l_{1},\cdots,l_{p}}(t_{1},\cdots,t_{p})=\\
&\delta_{l_1,0}\prod_{k=2}^p(1-\delta_{l_k,0}) C_{l_1,\cdots,l_{p}}(t_{1},\cdots,t_{p})\;.\nn
\end{align}
The values of the strictly positive rates $\mu_{j}(l_2,\cdots,l_p;\pi)>0$ for $2\leq j\leq p$ depend on the specific partition $\pi$ and read, supposing that $j\in B_{i}$,
\begin{align}
&\mu_j=\lambda_{l_j}+\lambda_{l_{j+1}}(1-\delta_{l_j,0})\\
&+\sum_{r=j+1}^{p}\lambda_{l_{r}}\prod_{m\neq i}\left[\delta_{r\in B_m}\Theta(j-\max\{s\in B_m,s<r\})\right]\;,\nn
\end{align}
with $\delta_{r\in B_m}=\sum_{s\in B_m}\delta_{r,s}$.
The values of $\lambda_r$'s for $r>j$ are included in $\mu_j$ if the times $t_r$ and $t_j$ do not belong to the same block and $t_j$ is larger than all the times $t_s<t_r$ belonging to the same block as $r$. 
 In particular, Eq. \eqref{con_correl} shows
that the $p$-time connected correlation decays exponentially with the associated rate $\mu_j$ as soon as one of the time-difference $\delta t_j=t_j-t_{j-1}$ becomes large.
\subsection{Proof of the general result}

We are now able to prove our main claim by computing the general expression of ${\cal Q}_{p}^{\alpha}(\omega_c)$. Changing variables in the integrals from the times 
$t_1,\cdots,t_p$ into $\delta t_2=t_2-t_1,\delta t_3,\cdots,\delta t_p=t_p-t_{p-1}$ and $t_p$, we can rewrite equation \eqref{cum_q_1} as
\begin{align}
&\moy{q_{\omega_c;\alpha}^p(t)}_{\rm co}=\label{q_exp}\\
&p!\, \omega_c^p \int_0^{t}dt_p\int_0^{t_p}d\delta t_{p}\cdots \int_0^{t_p-\sum_{k=3}^{p}\delta t_{k}}d\delta t_2 \nn\\
&\moy{\prod_{j=0}^{p-1} x_{\alpha}\left(t_p-\sum_{k=0}^{j-1}\delta t_{p-k}\right)}_{\rm co}\nn\\
&\prod_{j=0}^{p-1}\sin\left(\omega_c \left(t_p-\sum_{k=0}^{j-1}\delta t_{p-k}\right)\right)\;.\nn
\end{align}
It is clear that the variable $t_p$ is of order $O(t)$ in the large $t$ limit. The $p$-times
connected correlation term in the integrand $\moy{\prod_{j=1}^{p} x_{\alpha}\left(t_j\right)}_{\rm co}=\moy{\prod_{j=0}^{p-1} x_{\alpha}\left(t_p-\sum_{k=0}^{j-1}\delta t_{p-k}\right)}_{\rm co}$ is given in Eq. \eqref{con_correl}. This function does not decay with $t_p$ for fixed values of the $\delta t_j$'s but rather is a periodic function of this variable. On the other hand, we have seen that this function decays exponentially with all the variables $\delta t_{j}$'s  for $2\leq j\leq p$. We thus expect that the dominating contribution to the integral will come from $\delta t_j=O(1)$ while $t_p=O(t)$ in the large $t$ limit and we may safely replace the upper bounds $t_p-\sum_{k=j+1}^{p}\delta t_{k}$ in the integral over $\delta t_j$ by $+\infty$ in this limit. 

Introducing the Fourier expansion for the coefficients 
\be
C_{l_1,\cdots,l_p}(t_1,\cdots,t_p)=\sum_{k_1,\cdots,k_p=-\infty}^{\infty}C_{l_1,\cdots,l_p}^{k_1,\cdots,k_p}e^{i\sum_{j=1}^n k_j \omega_d t}\;,
\ee
as well as the identity
\be
\sin(\omega_c t)=\sum_{\sigma=\pm 1}\frac{(-\sigma)}{2i}e^{-i\sigma \omega_c t}\;,
\ee
one can express equation \eqref{q_exp} in the long time limit as
\begin{widetext}
\begin{align}
   \moy{q_{\omega_c;\alpha}^p(t)}_{\rm co}\approx &  p!\, \omega_c^p \sum_{k=1}^{p} (k-1)!(-1)^{k+1}\sum_{\pi \in P_k(p)}\prod_{B\in \pi}\sum_{l_1,\cdots,l_{p}=0}^{\infty}(1-\delta_{l_p,0})\delta_{l_1,0}\sum_{k_1,\cdots,k_{p}=-\infty}^{\infty}\sum_{\sigma_1,\cdots,\sigma_{p}=\pm 1}\prod_{m=1}^{p}\left(\frac{-\sigma_m}{2i}\right)\\
   &\times C_{l_1,\cdots,l_p}^{k_1,\cdots,k_p} \int_0^{t}dt_p \int_0^{\infty}d\delta t_p\cdots \int_0^{\infty}d\delta t_2 \prod_{j=2}^p e^{-\mu_j(l_2,\cdots,l_p;\pi)\delta t_j}\prod_{m=1}^{p} e^{i(k_m \omega_d-\omega_c \sigma_m) \left[t_p-\sum_{n=0}^{m-1}\delta t_{p-n}\right]}\;.\nn
\end{align}
\end{widetext}
In particular, for fixed values of $\{k_1,\cdots,k_p\}$ and $\{\sigma_1,\cdots,\sigma_p\}$, one can compute explicitly the integral over the variable $t_p$, which is the only term still depending on $t$ in the limit $t\gg \lambda_1^{-1}$. It yields
\begin{align}
\int_0^t dt_p\; e^{i\Omega_p t_p}&=e^{i\frac{\Omega_p t}{2}}\frac{2}{\Omega_p}\sin\left(\frac{\Omega_p t}{2}\right)\;,\\
\Omega_p&=\omega_d\sum_{j=1}^p k_j-\omega_c \sum_{j=1}^p \sigma_j\;.\label{Eq.ResCondition}
\end{align}
In particular, it is only of order $t$ in the special case where $\Omega_p=0$, giving a resonance condition for each of the cumulants. The only terms that contribute to $\mathcal{Q}_p(\omega_c)$, are by Eq.(\ref{Eq:Q_p^alpha}) these specific resonant contributions which scale as $O(t)$, namely for which the right hand side of Eq.(\ref{Eq.ResCondition}) vanishes. There are several important consequences for this observation. Let us therefore focus on the equation
\begin{align}
    \omega_d\sum_{j=1}^p k_j-\omega_c \sum_{j=1}^p \sigma_j=0\;.\label{Eq.ResCondition1}
\end{align}
This equality holds when both sums are zero independently
\begin{align}
    \sum_{j=1}^p k_j &=0\nonumber\\
    \sum_{j=1}^p \sigma_j&=0 \;\label{Eq.ResCondition2}
\end{align}
or when $\omega_c$ and $\omega_d$ achieve some commensurate values
\begin{align}
    \frac{\omega_c}{\omega_d}=\frac{\sum_{j=1}^p k_j}{\sum_{j=1}^p \sigma_j} \;.\label{Eq.ResCondition3}
\end{align}
For incommensurate $\omega_d$ and $\omega_c$, Eq.(\ref{Eq.ResCondition3}) cannot hold, since both the numerator and denominator in the right hand side of the equation are integers. Therefore, the only valid resonance condition corresponds to the case where both quantities in Eq.(\ref{Eq.ResCondition2}) are zero. 
However, the second line of \eqref{Eq.ResCondition2} cannot hold for odd $p$, since $\sigma_j=\pm 1$ and in particular the second line of Eq. \eqref{Eq.ResCondition2} cannot vanish. Thus, for incommensurate frequency ratios all the odd cumulants are exactly zero. For even values of $p$, it is possible that each of the sums above is zero separately, and therefore generically even cumulants do not vanish.

Next, let us consider commensurate frequencies: in this case, there are two options for condition Eq.(\ref{Eq.ResCondition}) to hold: either Eq.(\ref{Eq.ResCondition2}) holds, or instead Eq.(\ref{Eq.ResCondition3}) can hold. For even values of $p$, the latter case is possible whenever the relations between $\omega_c$ and $\omega_d$ is of the form  
\be
\omega_c=\frac{n}{2z}\omega_d\;,\;\;n\in\mathbb{N}^*\;,\;\;1\leq z\in\mathbb{N}\leq \frac{p}{2}\;,\label{Eq.Even_p_Spikes}
\ee
since $\sum_j k_j$ can take any integer value, whereas $\sum_j \sigma_j$ is bounded between $-p$ and $p$. Similarly, for odd values of $p$, condition Eq.(\ref{Eq.ResCondition}) can hold when
\be
\omega_c=\frac{n}{2z+1}\omega_d\;,\;\;n\in\mathbb{N}^*\;,\;\;0\leq z\in\mathbb{N}\leq \frac{p-1}{2}\;.\label{Eq.Odd_p_Spikes}
\ee
In both the even and odd cases we expect contributions to the cumulants in addition to those that appear in the non-commensurate cases. Therefore, generically they have discontinuities at these specific commensurate frequencies. 

At this point, one can rightfully suspect that the cumulant generating function $\chi_{\mu}(\omega_c)$, which holds the discontinuity of every cumulant, is for fixed $\mu$ an everywhere discontinuous function of the alternating charge's frequency $\omega_c$. However, we next show that it is a continuous function over the frequencies $\omega_c$ incommensurate with $\omega_d$, similarly to the Thomae's function \cite{thomaes}. To this end, we next examine the structure of ${\cal Q}_{p}^{\alpha}(\omega_c)$ as a function of $p$.

\subsection{General expression of ${\cal Q}_{p}^{\alpha}(\omega_c)$}
Using the tools presented above it is possible to obtain the general expression for the cumulant of arbitrary order $p$. It reads
\begin{widetext}
\begin{align}
    {\cal Q}_{p}^{\alpha}(\omega_c)=&p!\sum_{k=1}^{p} (k-1)!(-1)^{k+1}\sum_{\pi \in P_k(p)}\prod_{B\in \pi}\sum_{k_1,\cdots,k_p=-\infty}^{\infty}\sum_{\sigma_1,\cdots,\sigma_p=\pm 1}\sum_{l_1,\cdots,l_{p}=0}^{\infty}(1-\delta_{l_p,0})\delta_{l_1,0}\label{Q_p_gen}\\
&\times\prod_{j=1}^p \left(\frac{-\sigma_j}{2i}\right)\frac{\displaystyle\prod_{i=1}^k\left[\delta_{l_{B_i(1)},0}\prod_{k=2}^{n_i}(1-\delta_{l_{B_i(k)},0}) C_{l_{B_i(1)},\cdots,l_{B_i(n_i)}}^{k_{B_i(1)},\cdots,k_{B_i(n_i)}}\right]}{\displaystyle \prod_{j=2}^p\left[\mu_j(l_2,\cdots,l_p;\pi)+i\sum_{l=j}^p\left(\sigma_l\omega_c-k_l\omega_d\right)\right]}\omega_c^p\, \delta_{\sum_{j=1}^p \left(\omega_d k_j-\omega_c \sigma_j\right),0}\;.\nn
\end{align}
\end{widetext}
The Kronecker delta function in the second line of this expression ensures that the resonance condition is fulfilled. For incommensurate frequencies, this happens only when Eq. \eqref{Eq.ResCondition2} holds but for commensurate ratio, this also happens when Eq. \eqref{Eq.ResCondition3} holds. 

We now explore, for a fixed $\mu$, the continuity of the cumulant generating function
\be
\chi_{\mu}(\omega_c)=\sum_{p=1}^{\infty}\frac{\mu^p}{p!}{\cal Q}_{p}^{\alpha}(\omega_c)\;.
\ee
As discussed above, each term of this series has specific discontinuities as detailed in Eqs. \eqref{Eq.Even_p_Spikes} and \eqref{Eq.Odd_p_Spikes}. 
For a sequence of rational ratios that convergence to an irrational ratio, namely a sequence of $(n_i,m_i)$ and a corresponding $\omega_c^i$
\be
\omega_c^i=\omega_d\frac{n_i}{m_i}\;,
\ee
such that $a=\lim_{i\to \infty} n_i/m_i$ is an irrational number, the resonance condition in Eq. \eqref{Eq.ResCondition3} imposes that the cumulants which display a discontinuity must be of increasing order $p_i$. To this end, let us examine carefully the denominator of equation \eqref{Q_p_gen}. For any a value of $p_i$, the resonance condition imposes
\be
\frac{\sum_{j=1}^{p_i} k_j}{\sum_{j=1}^{p_i} \sigma_j}=\frac{n_i}{m_i}\;.
\ee
For a random strings $\{\sigma_j, j=1,\cdots,p\}$ and $\{k_j, j=1,\cdots,p\}$ with the constraint of fulfilling the latter equation, the resonance condition imposes the following equality:
\be
\left|\sum_{l=j}^{p_i}\left(\sigma_l\omega_c-k_l\omega_d\right)\right|=\left|\sum_{l=1}^{j-1}\left(\sigma_l\omega_c-k_l\omega_d\right)\right|\;.
\ee
The left hand side of this equation scales, in the large $j$ limit, as $O(\sqrt{j})$. Thus we expect the product appearing in the denominator of Eq. \eqref{Q_p_gen}
\be
\prod_{j=2}^{p_i}\left[\mu_j+i\sum_{l=j}^{p_i}\left(\sigma_l\omega_c-k_l\omega_d\right)\right]
\ee
to grow in modulus rather rapidly as a function of $p_i$ (as we expect $\mu_j=O(1)$, this term will be of order $\sqrt{p_i!}$). Therefore we expect that ${\cal Q}_{p_i}^{\alpha}(\omega_c)/(p_i!)$ goes to zero as $p_i\to \infty$. The discontinuous part of the CGF at frequency $\omega_c^i$ are provided by the terms ${\cal Q}_{p}^{\alpha}(\omega_c)/(p!)$ of order $p>p_i$ and these rescaled cumulants are getting smaller and smaller as $p_i$ is increased, we expect that in the limit where $i\to \infty$, the discontinuities vanish completely from the CGF.

To get some insight on equation \eqref{Q_p_gen}, we next use it to extract the explicit expressions for the first two cumulants, which are the main objects of interest in most cases. 

\subsection{Average alternating current}

Rather than simply plugging $p=1$ in Eq. \eqref{Q_p_gen}, it is both easy and insightful to rapidly re-derive the result in this particular case at finite but large time $t$. To this end, we first compute the finite time first cumulant $\moy{q_{\omega_c;\alpha}(t)}$. It reads
\be
\moy{q_{\omega_c;\alpha}(t)}=\omega_c\sum_{l_1=0}^{\infty}\int_0^{t}dt_1 \sin(\omega_c t_1)e^{-\lambda_1 t_1}C_{l_1}(t_1)\;,  \label{q_wc_1}
\ee
where we have used the Floquet expansion of the propagator, to obtain
\be
\moy{x_{\alpha}(t)}=\sum_{l_1=0}^{\infty}e^{-\lambda_1 t}C_{l_1}(t)\;,
\ee
and we remind that the function $C_{l_1}(t)$ is a periodic function of $t$ with fundamental frequency $\omega_d$.
We introduce the Fourier expansion
\begin{align}
C_{l_1}(t)&=\sum_{k=-\infty}^{\infty}C_{l_1}^{k} e^{i k\omega_d t} \;. 
\end{align}
Note that $\moy{x_{\alpha}(t)}$ is a real function of $t$ such that if the Floquet eigenvalue $\lambda_{l_1}$ is real, the associated function $C_{l_1}(t)$ must also be real and $C_{l_1}^{-n}={C_{l_1}^{n}}^*$ (we denote by $z^*$ here and in the following the complex conjugate of $z$). If the Floquet eigenvalue $\lambda_{l_1}$ is complex, it will come in a complex conjugate pair, say $\lambda_{l_1}^*=\lambda_{l_1+1}$ and one must have instead that $C_{l_1}(t)=C_{l_1+1}(t)^*$, which yields $C_{l_1}^{-n}={C_{l_1+1}^{n}}^*$. Inserting this expansion into Eq. \eqref{q_wc_1}, we obtain in the regime where $t\gg \lambda_1^{-1}$
\begin{widetext}
\begin{align}
    \moy{q_{\omega_c;\alpha}(t)}\approx&\sum_{k=-\infty}^{\infty}\frac{\omega_c\,C_{0}^{k}}{\omega_c^2-k^2\omega_d^2}\left[\omega_c(1-e^{i k\omega_d t}\cos(\omega_c t))+i k\omega_d e^{i k\omega_d t}\sin(\omega_c t)\right]+\sum_{k=-\infty}^{\infty}\sum_{l_1=1}^{\infty} \frac{\omega_c^2\,C_{l_1}^{k}}{\omega_c^2+(\lambda_{l_1}-i k\omega_d)^2}\;.
\end{align}
\end{widetext}
Using that ${C_{0}^{k}}^*=C_{0}^{-k}$ since $\lambda_0=0$, in the double scaling limit $t\to \infty$, $\omega_c-n\omega_d\to 0$ with $\phi=(\omega_c-n\omega_d)t=O(1)$ fixed, the rescaled alternating charge takes the following scaling form
\begin{align}
\frac{\moy{q_{\omega_c;\alpha}(t)}}{t}&\approx -\omega_c\, {\cal J}_n((\omega_c-n\omega_d)t)\label{Q_1_t}\\
{\cal J}_n(\phi)&=-\Re\left[ C_{0}^{n}\right]\frac{2\sin^2\left(\phi/2\right)}{\phi}+\Im\left[ C_{0}^{n}\right]\frac{\sin\left(\phi\right)}{\phi}\;.\nn
\end{align}

We are mainly interested in the limit ${\cal Q}_{1}^{\alpha}(\omega_c)=\lim_{t\to \infty}\moy{q_{\omega_c;\alpha}(t)}/t$, which can simply be obtained from this result as
\begin{align}
{\cal Q}_{1}^{\alpha}(\omega_c)&=-\omega_c\sum_{n=0}^{\infty}{\cal J}_n(0)\delta_{n\omega_d,\omega_c}\nn\\
&=-\omega_c\sum_{n=0}^{\infty}\Im\left[ C_{0}^{n}\right] \delta_{\omega_c,n\omega_d}\label{Q_1}\;.
\end{align}
This structure is clearly discontinuous in the $t\to \infty$ limit as any term for $\omega_c\neq n\omega_d $ decays as $1/t$.
The average alternating charge ${\cal Q}_1^{\alpha}(\omega_c)$ is zero for any frequency $\omega_c\neq n\omega_d$ that is not a harmonic of the driving frequency $\omega_d$ as one could have naively guessed: there is no average probability current in periodically driven systems, except at frequencies which are an integer multiple of the driving frequency.

\subsection{Variance of the alternating charge}

We next turn to the computation of the variance of the alternating charge. In order to compute this object, we first compute the connected two-time correlation function for $t_2>t_1\gg \lambda_1^{-1}$
\begin{align}
    \moy{x_\alpha(t_1)x_\alpha(t_2)}_{\rm co}\approx \sum_{l_2=1}^{\infty}e^{-\lambda_{l_2}(t_2-t_1)}C_{0,l_2}(t_1,t_2)\;,
\end{align}
where we have used that $C_{l_1,l_2=0}(t_1,t_2)=C_{l_1}(t_1)C_{0}(t_2)$. The expression of the second cumulant is again particularly simple as only the identity partition of $\{1,2\}$ yields a non-zero contribution in the connected $2$-times correlation function. It reads
\begin{align}
&{\cal Q}_{2}^{\alpha}(\omega_c)=\label{Q_2}\\
&\sum_{k_1,k_2=-\infty}^{\infty}\sum_{\sigma_1,\sigma_2=\pm 1}\sum_{l_2=1}^{\infty}\frac{-\omega_c^2\,\sigma_1\sigma_2\,C_{0,l_2}^{k_1,k_2}\,\delta_{(k_1+k_2)\omega_d,(\sigma_1+\sigma_2)\omega_c}}{2[\lambda_{l_2}+i(\sigma_2\omega_c-k_2\omega_d)]}\nn\\
&={\cal Q}_{2}^{{\rm b},\alpha}(\omega_c)-\sum_{n=1}^{\infty}\sum_{l_2=1}^{\infty}\sum_{k=-\infty}^{\infty}\sum_{\sigma=\pm 1} \frac{\omega_c^2 \,C_{0,l_2}^{\sigma n-k,k}\,\delta_{2\omega_c,n\omega_d} }{2[\lambda_{l_2}+i(\sigma\omega_c-k\omega_d)]}\,,\nn
\end{align}
where the continuous background ${\cal Q}_{2}^{{\rm b},\alpha}(\omega_c)$ reads
\be
{\cal Q}_{2}^{{\rm b},\alpha}(\omega_c)=\sum_{l_2=1}^{\infty}\sum_{k=-\infty}^{\infty} \frac{\omega_c^2(\lambda_{l_2}-i k\omega_d)\,C_{0,l_2}^{-k,k}}{(\lambda_{l_2}-i k\omega_d)^2+\omega_c^2}\,. 
\ee
In contrast to ${\cal Q}_{1}^{\alpha}(\omega_c)$, which is zero except for frequencies that satisfy $\omega_c=n\omega_d$, the rescaled variance ${\cal Q}_{2}^{\alpha}(\omega_c)$ is generically non-zero for any frequency $\omega_c$ and equal to ${\cal Q}_{2}^{{\rm b},\alpha}(\omega_c)$ away from the integer and half-integer multiple of $\omega_d$. Therefore, there are current fluctuations even at frequency with no average current, as expected.

It is instructive to compare these results with the power spectral density calculated for the stochastic resonance setup \cite{shneidman1994power,nikitin2007asymmetric,nikitin2005asymmetry}. There, the peaks appear only at integer multiples of $\omega_d$, whereas in our setup they appear at both integer and half integer multiples of $\omega_d$.

The expressions for higher order cumulants can be obtained more explicitly but are quite cumbersome. In Appendices \ref{third} and \ref{fourth}, we respectively give explicit expressions for the third and fourth cumulant of the alternating charge.


\subsection{Large $\omega_c$ limit of the CGF}

In the case $\omega_c\gg \omega_d$, a simplified expression for the tilted operator ${\cal L}_{\mu}({\bf x},\nabla_{\bf x},t)$ can be obtained. The bare dynamics that evolves over the period $T_d$ is not fast enough to change over a period $T_c\ll T_d$, such that the terms depending on $\omega_c$ can be replaced in practice by their cycle average value over the period $T_c$, i.e.
\begin{align}
&{\cal L}_{\mu}({\bf x},\nabla_{\bf x},t)G_{\mu}({\bf x},t|{\bf x}_0)\approx\overline{{\cal L}_{\mu}({\bf x},\nabla_{\bf x},t)}^{\omega_c}G_{\mu}({\bf x},t|{\bf x}_0)\;,\nn\\
&\overline{{\cal L}_{\mu}({\bf x},\nabla_{\bf x},t)}^{\omega_c}={\cal L}({\bf x},\nabla_{\bf x},t)+\frac{\mu^2}{2}D_{\alpha\alpha}({\bf x},t)\;,
\end{align}
 where $\overline{A}^{\omega_c}=T_c^{-1}\int_0^{T_c} A(t)dt$. In the specific case where $D_{\alpha\alpha}({\bf x},t)=D_{\alpha\alpha}(t)$ is independent of ${\bf x}$, one can show that the tilted propagator can be simply expressed as a function of the bare propagator as
\be
G_{\mu}({\bf x},t)\approx G({\bf x},t|{\bf x}_0,0)e^{\frac{\mu^2}{2}\int_0^t D_{\alpha\alpha}(\tau)d\tau}\;.
\ee
Form this result, one obtains that as $\omega_c\to\infty$
\begin{align}
\chi_{\mu}(\omega_c)&=\lim_{t\to \infty}\frac{1}{t}\ln \int d{\bf x}\, G_{\mu}({\bf x},t)\approx\frac{\mu^2}{2}\overline{D_{\alpha\alpha}}^{\omega_d}\;,\\
{\cal Q}_{p}(\omega_c)&=\left.\partial_\mu^p\chi_{\mu}(\omega_c)\right|_{\mu=0}\approx\delta_{p,2}\overline{D_{\alpha\alpha}}^{\omega_d}\;,
\end{align}
where $\overline{D_{\alpha\alpha}}^{\omega_d}=T_d^{-1}\int_0^{T_d} D_{\alpha\alpha}(\tau)d\tau$. In particular, it implies that both the typical and atypical fluctuations of the alternating charge are Gaussian in the $\omega_c\to \infty$ limit.

\section{Specific examples} \label{examples}

We have derived general expressions for the first two cumulants of the alternating charge. Next, we demonstrate these results in specific examples, which are commonly used in statistical mechanics and where a full analytical computation is possible.

\subsection{Underdamped Ornstein Uhlenbeck}

As a first example, we  consider the following dynamics
\begin{align}
    \dot v(t)&=\dot\mu(t)-\Gamma(t)\left[v(t)-\mu(t)+\Omega(t)x(t)-\sqrt{2D(t)}\eta(t)\right]\;,\nn\\
    \dot x(t)&=v(t)\;,
\end{align}
where $\eta(t)$ is a Gaussian white noise with zero mean $\moy{\eta(t)}=0$ and variance $\moy{\eta(t_1)\eta(t_2)}=\delta(t_1-t_2)$. The damping $\Gamma(t)$, trapping frequency $\Omega(t)$, diffusion coefficient $D(t)$ and drift $\mu(t)$ are all time-periodic functions with fundamental angular frequency $\omega_d$. Keeping track of both $(x,v)$ at time $t$, the system is Markovian and the general theory described in the sections above applies to this case. 

By linearity of the above equation, both the position $x(t)$ and the speed $v(t)$ are expressed as (infinite) linear combinations of the Gaussian random variables $\eta(\tau)$ for $\tau\in[0,t]$ and thus also have Gaussian fluctuations. The same reasoning applies to the empirical
alternating charge $Q_{\omega_c; x}(t)$, which is expressed as (infinite) sums of $x(\tau)$ for $\tau\in[0,t]$. It yields that all the cumulants beyond the variance are zero, i.e.
\begin{align}
&{\cal Q}_p^{x}(\omega_c)=0\;,\;\;p>2\;,\\
&\chi_{\mu}(\omega_c)=\mu {\cal Q}_1^{x}(\omega_c)+\frac{\mu^2}{2}{\cal Q}_2^{x}(\omega_c)\;.
\end{align}
A direct consequence is that the large deviation function takes the quadratic form
\be
\Phi_{\omega_c}(J)=\frac{(J-{\cal Q}_1^{x}(\omega_c))^2}{2{\cal Q}_2^{x}(\omega_c)}\;.
\ee
The joint bare propagator $G(v,x,t|v_0,x_0,t_0)$ of the speed $v$ and position $x$ satisfies the equation
\begin{align}
\partial_t G=&-(v+\mu(t))\partial_x G+\Gamma(t)\partial_v\left[(v+\Omega(t)x) G\right]\label{FP_x_v}\\
&+\Gamma(t)D(t)\partial_v^2 G\;.\nn
\end{align}
We consider a simple example with constant damping $\Gamma(t)=\Gamma_0$ and trapping frequency $\Omega(t)=\Omega_0$ and introduce the Fourier series
\be
\mu(t)=\sum_{k=-\infty}^{\infty}\mu_k\, e^{ik\omega_d t}\;,\;\;D(t)=\sum_{k=-\infty}^{\infty}D_k\, e^{ik\omega_d t}\;,
\ee
which correspond to periodic forcing of the drift and the diffusion coefficient.
We suppose that the functions $\mu(t)$ and $D(t)$ are real, such that for all $k$, $\mu_k=\mu_{R,|k|}+i\,\sign{k} \mu_{I,|k|}$ and similarly $D_k=D_{R,|k|}+i\,\sign{k} D_{I,|k|}$. In this case, the position at time $t$, starting with the initial condition $(x(0),v(0))=(0,0)$ is given by
\begin{align}
x(t)=&\sqrt{\frac{\Gamma_0}{\Gamma_0-4\Omega_0}}\sum_{\alpha=\pm 1}\alpha\\
&\times\int_0^{t}d\tau\, e^{-\frac{(t-\tau)}{2}\left[\Gamma_0-\alpha\sqrt{\Gamma_0^2-4\Gamma_0\Omega_0}\right]}\,\left[\eta(\tau)+\mu(\tau)\right]\;.\nn
\end{align}
This expression allows to identify the Floquet spectrum $\lambda_0=0\leq \lambda_1\leq \lambda_2$ and $\lambda_k=+\infty$ for $k\geq 3$ with
\begin{align}
    \lambda_1&=\frac{\Gamma_0}{2}-\frac{\sqrt{\Gamma_0^2-4\Gamma_0 \Omega_0}}{2}\;,\nn\\ \lambda_2&=\frac{\Gamma_0}{2}+\frac{\sqrt{\Gamma_0^2-4\Gamma_0 \Omega_0}}{2}\;.\label{spectrum_BM}
\end{align}
Using the expression of the position, one can show that
\begin{align}
\moy{x(t)}=&\sqrt{\frac{4\Gamma_0}{\Gamma_0-4\Omega_0}}\\
&\times\int_0^{t}d\tau\, e^{-\frac{\Gamma_0}{2}\tau}\sinh\left(\sqrt{\Gamma_0^2-4\Gamma_0\Omega_0}\frac{\tau}{2}\right)\,\mu(t-\tau)\;.\nn
\end{align}
Inserting the Fourier series of $\mu(t)$ and taking the long-time limit $t\gg \lambda_1^{-1}$, one can identify this expression with
\begin{align}
\moy{x(t)}\approx& \sum_{k=-\infty}^{\infty}C_0^k e^{i k t}\;,
\end{align}
which allows to obtain the expression of the coefficients
\be
C_0^{k}=\frac{\Gamma_0\mu_k}{\Gamma_0(\Omega_0+ik\omega_d)-k^2\omega_d^2}\;.\label{C_0_k}
\ee
Using this expression, we obtain the expression of the first cumulant
\begin{align}
{\cal Q}_1^{x}(\omega_c)&=\sum_{n=1}^{\infty}Q_{1,n}\delta_{n\omega_d,\omega_c}\;,\label{Q_1_BM}\\
Q_{1,n}=&\frac{\mu_{R,n}\Gamma_0^2n^2\omega_d^2+\mu_{I,n}\Gamma_0 n\omega_d(n^2\omega_d^2-\Gamma_0\Omega_0)}{\Gamma_0^2\Omega_0^2+\Gamma_0(\Gamma_0-2\Omega_0)n^2\omega_d^2+n^4\omega_d^4}\;.\nn
\end{align}

\begin{figure}[h]
	\centering
	\includegraphics[width=0.9\linewidth]{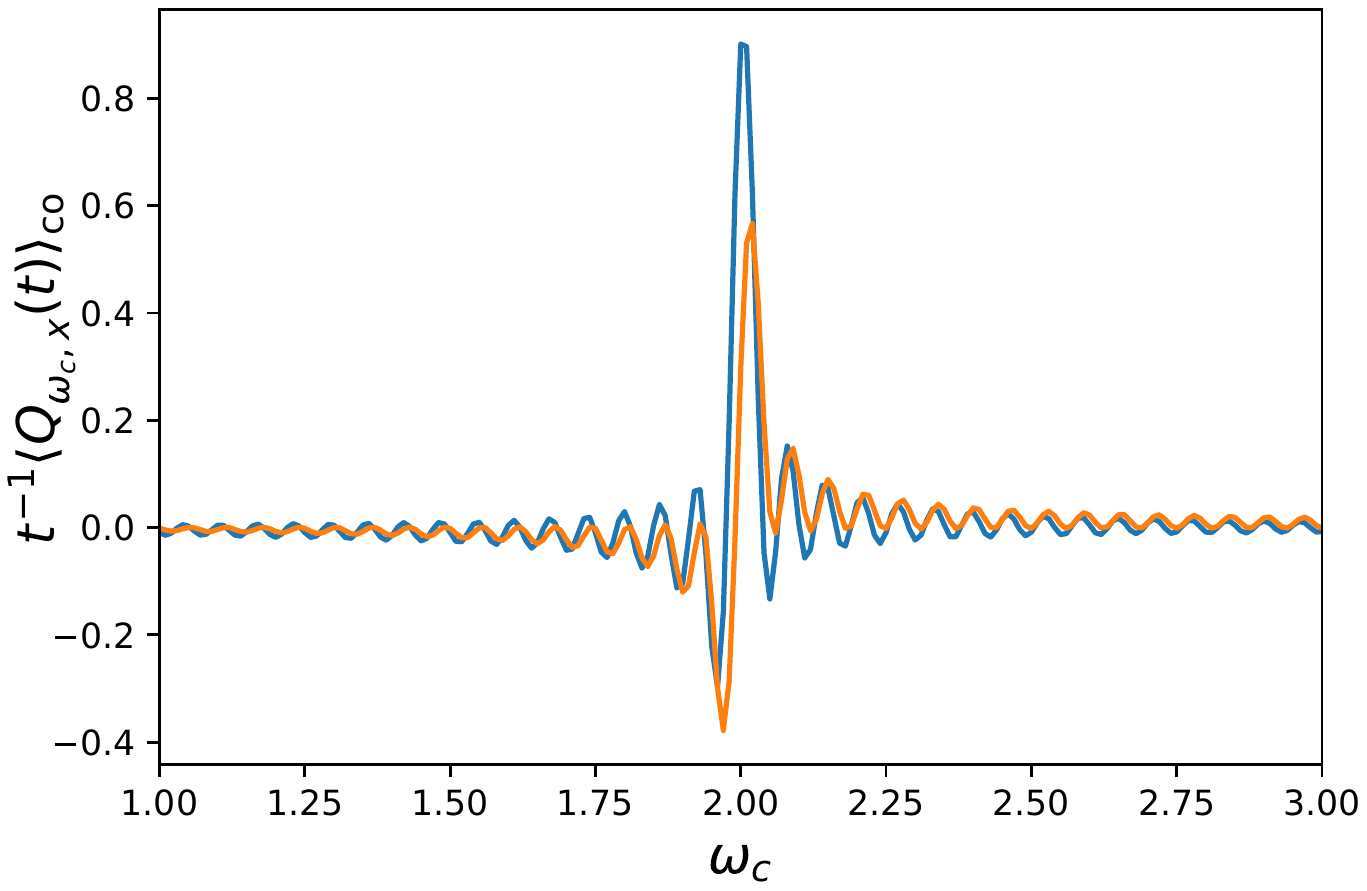}
	\caption{Comparison between the time-averaged alternating current $\moy{Q_{\omega_c}(t)}/t$ (in blue) computed numerically for a measurement time $t=10^2$ for a periodically driven underdamped Brownian particle with $\Gamma_0=2$, $\Omega_0=1$, $\omega_d=1$ and for $\mu(t)=3/2\cos(2\omega_d t)$ and the analytical expression at finite time (in orange) given by inserting the coefficient $C_0^{k}$ given in \eqref{C_0_k} into Eq. \eqref{Q_1_t}.}
	\label{Fig:cum_1_BM}
\end{figure}

One can similarly compute the two-time correlation function
\begin{align}
&\moy{x(t_1)x(t_2)}_{\rm co}=\moy{x(t_1)x(t_2)}-\moy{x(t_1)}\moy{x(t_2)}\\
&=\sum_{\alpha_1,\alpha_2=\pm 1}\frac{2\Gamma_0\alpha_1\alpha_2}{\Gamma_0-4\Omega_0}\nn e^{-\frac{(t_2-t_1)}{2}\left[\Gamma_0-\alpha_2\sqrt{\Gamma_0^2-4\Gamma_0\Omega_0}\right]}\nn\\
&\times\int_0^{t_1}d\tau\,e^{-\frac{(t_1-\tau)}{2}\left[2\Gamma_0-(\alpha_1+\alpha_2)\sqrt{\Gamma_0^2-4\Gamma_0\Omega_0}\right]}D(\tau)\;.\nn
\end{align}
From this expression, it is possible to identify the coefficients 
\be
C_{0,l_2}^{k_1,k_2}=\frac{2\Gamma_0^{3/2}D_{k_1}}{\sqrt{\Gamma_0-4\Omega_0}}\frac{(\delta_{l_2,1}-\delta_{l_2,2})\delta_{k_2,0}}{[\Gamma_0+ik_1 \omega_d][2\lambda_{l_2}+ik_1 \omega_d]}\;.\label{C_2_BM}
\ee
One can check that for $\Gamma_0<4\Omega_0$, we have that $\Im[\lambda_1]=\Im[\lambda_2]=0$ and $\Im[C_{0,l_2}^{0,0}]=0$ while for $\Gamma_0>4\Omega_0$ we have $\lambda_1=\lambda_2^*$ such that $C_{0,1}^{0,0}={C_{0,2}^{0,0}}^*$. After simplifications, one can obtain explicitly the value of the rescaled variance of the alternating charge by inserting the expression of the coefficients \eqref{C_2_BM} and of the spectrum \eqref{spectrum_BM} into \eqref{Q_2}
\begin{align}
&{\cal Q}_2^{x}(\omega_c)=\frac{\Gamma_0^2 \omega_c^2 D_0}{\Gamma_0^2\Omega_0^2+\Gamma_0(\Gamma_0-2\Omega_0)\omega_c^2+\omega_c^4}\\
&-\sum_{n=1}^{\infty}\frac{\Gamma_0^2\omega_c^2 D_{R,n}(\Gamma_0^2\Omega_0^2-\Gamma_0(\Gamma_0+2\Omega_0)\omega_c^2+\omega_c^4)}{(\Gamma_0^2\Omega_0^2+\Gamma_0(\Gamma_0-2\Omega_0)\omega_c^2+\omega_c^4)^2}\delta_{n\omega_d,2\omega_c}\nn\\
&-\sum_{n=1}^{\infty}\frac{2\Gamma_0^3\omega_c^3D_{I,n}(\Gamma_0\Omega_0-\omega_c^2)}{(\Gamma_0^2\Omega_0^2+\Gamma_0(\Gamma_0-2\Omega_0)\omega_c^2+\omega_c^4)^2}\delta_{n\omega_d,2\omega_c}\;.\nn
\end{align}

\begin{figure}[h]
	\centering
	\includegraphics[width=0.9\linewidth]{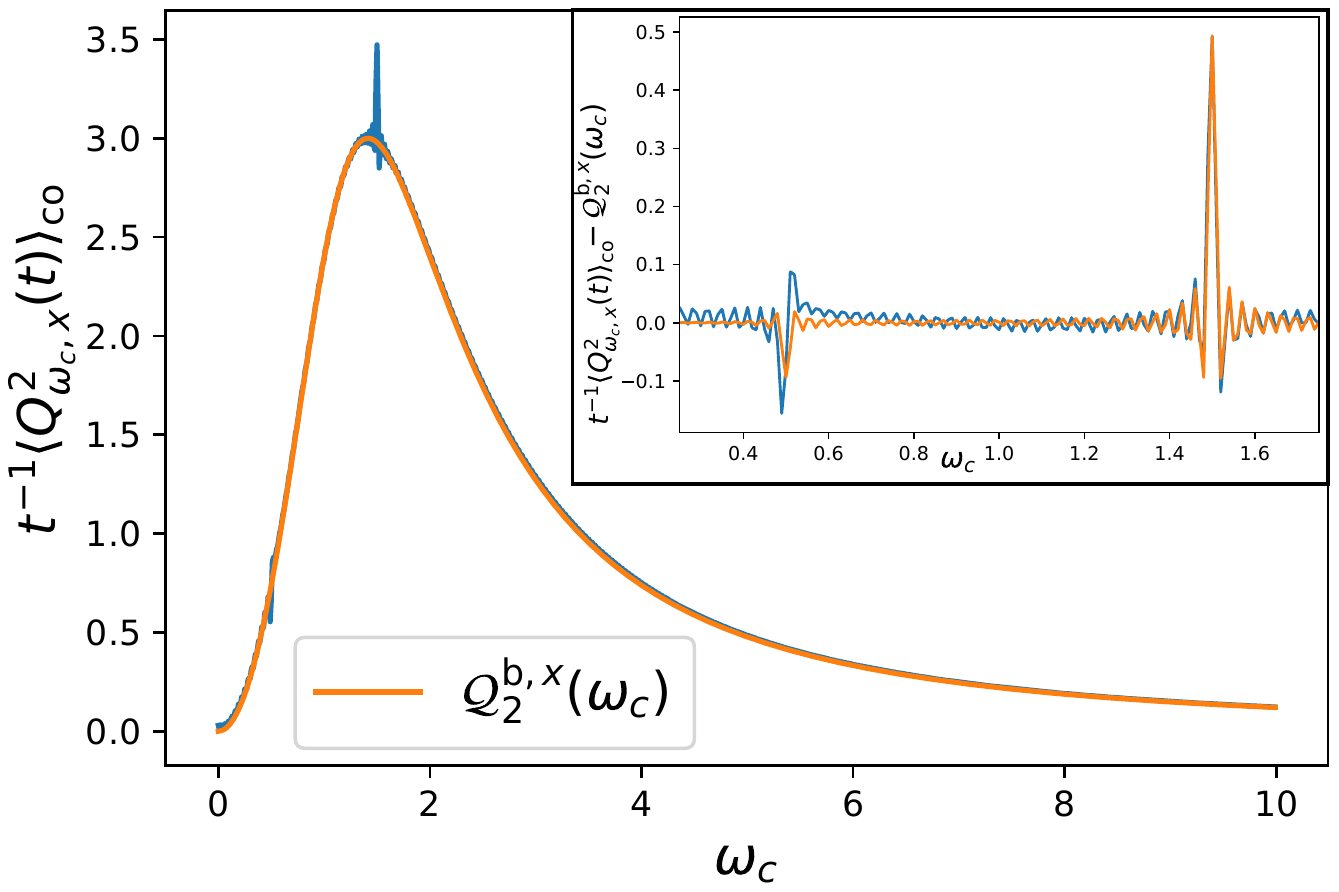}
	\caption{Comparison between the rescaled variance of the alternating charge $\moy{Q_{\omega_c,x}^2(t)}_{\rm co}/t$ (in blue) computed numerically for a measurement time $t=10^2$ for a periodically driven underdamped Brownian particle with $\Gamma_0=2$, $\Omega_0=1$, $\omega_d=1$ and for $D(t)=3(1+1/2\cos(\omega_d t)+1/3\cos(3\omega_d t))$ and the background value ${\cal Q}_2^{{\rm b},x}(\omega_c)$ (in orange) given in Eq. \eqref{var_BM}. The measurement time is $t=10^2$. In inset we have plotted (in blue) the difference between the numerical result and the analytical background, highlighting the  presence of spikes at values of $\omega_c$ corresponding to half the Fourier components of $D(t)$, i.e. for $\omega_c/\omega_d=1/2,3/2$. The orange line in inset corresponds to the finite time  analytical prediction for the spikes.}
	\label{Fig:cum_2_BM}
\end{figure}

Note that the continuous background,
\be
{\cal Q}_2^{{\rm b},x}(\omega_c)=\frac{\Gamma_0^2 \omega_c^2 D_0}{\Gamma_0^2\Omega_0^2+\Gamma_0(\Gamma_0-2\Omega_0)\omega_c^2+\omega_c^4}\;,\label{var_BM}
\ee
is identical to the variance of the fluctuating charge for a system with constant diffusion coefficient $D_0$, i.e. in the absence of any periodic drive. The only effect of the periodic drive is thus the emergence of the discontinuities at frequencies $\omega_c=n\omega_d/2$ for integer $n$. The continuous part of the spectrum of ${\cal Q}_2^{{\rm b},x}(\omega_c)$ presents a local maximum ${\cal Q}_2^{{\rm b},x}(\omega_c=\sqrt{\Gamma_0\Omega_0})=D_0$ for $\sqrt{\Gamma_0\Omega_0}$ away from integer and half-integer multiples of $\omega_d$ (unless $\omega_d=2\sqrt{\Gamma_0\Omega_0}/m$ for some $m\in \mathbb{N}^*$). In Fig. \ref{Fig:cum_2_BM}, we show a comparison between our analytical computation for the rescaled variance ${\cal Q}_2^{x}(\omega_c)$ as a function of $\omega_c$. The agreement is excellent for the background ${\cal Q}_2^{{\rm b}, x}(\omega_c)$ and relatively good for the spikes. 

While the average alternating charge ${\cal Q}_1^{x}(\omega_c)$ is independent of the diffusion coefficient $D(t)$, the variance ${\cal Q}_2^x(\omega_c)$ is independent of the drift $\mu(t)$. For a finite value of $\Gamma_0$, one has that ${\cal Q}_2^x(\omega_c)\to 0$ as $\omega_c\to \infty$ which is consistent with the absence of a term of the form $\partial_x^2 G$ in equation Eq. \eqref{FP_x_v}. Finally, the overdamped limit can easily be obtained in this example by taking the limit $\Gamma_0\to \infty$. In this limit, the evolution equation for the effective propagator $G(x,t|x_0,t_0)$ reads
\be
\partial_t G=\Omega(t)\partial_x\left((x-\mu(t))G\right)+D(t)\partial_x^2 G\;.
\ee
The background variance reads ${\cal Q}_2^{\rm b}(\omega_c)=D_0\omega_c^2/(\Omega_0^2+\omega_c^2)$.
In the limit $\omega_c\to \infty$, one has that ${\cal Q}_2(\omega_c)\to D_0$, which is consistant with the prefactor of $\partial_x^2 G$ being $D(t)$ and $\overline{D(t)}^{\omega_d}=D_0$.

\subsection{Periodically driven run-and-tumble particle}

\begin{figure}[h]
	\centering
	\includegraphics[width=0.9\linewidth]{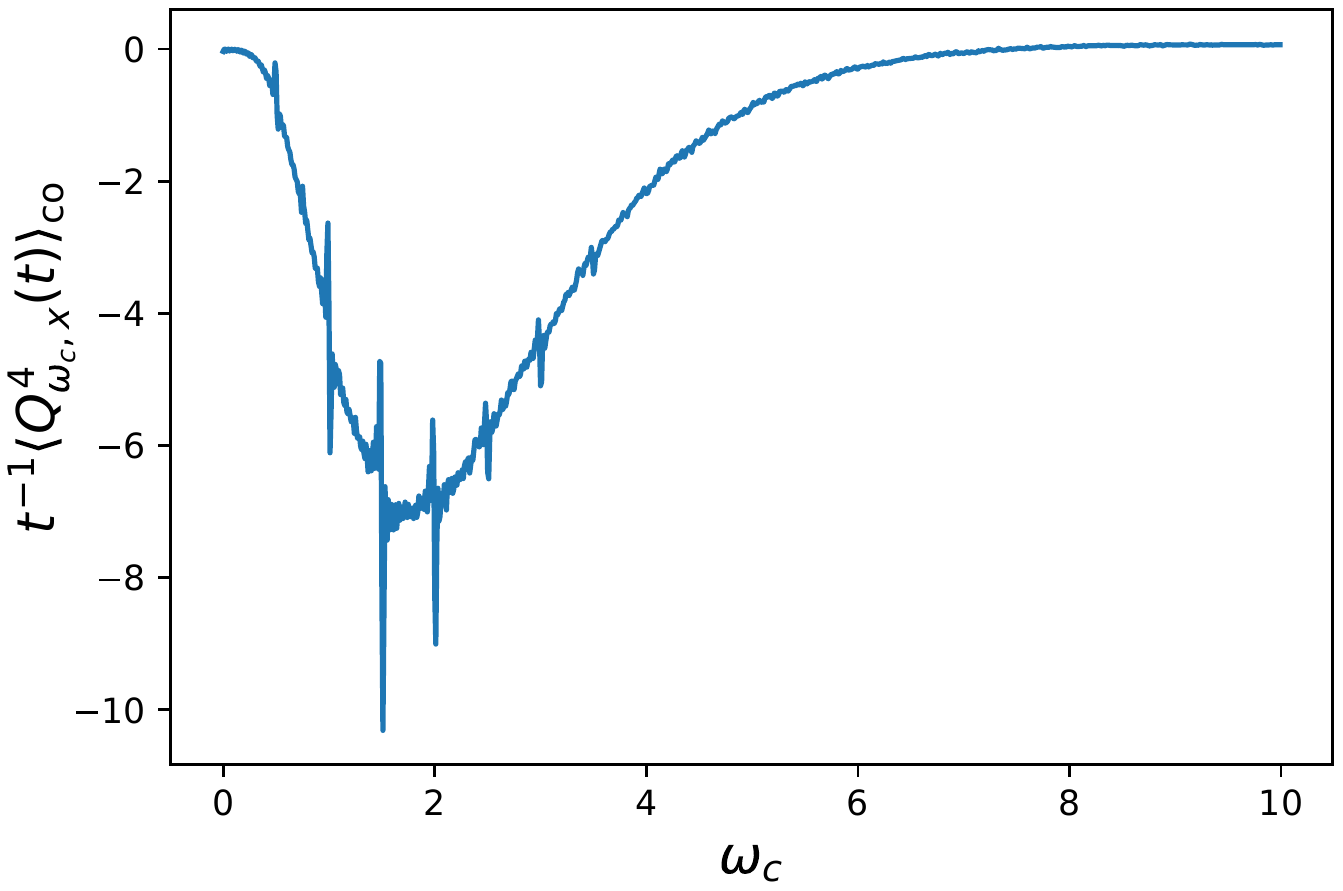}
	\caption{Plot of the rescaled fourth order cumulant of the alternating charge $\moy{Q_{\omega_c,x}^4(t)}_{\rm co}/t$ (in blue) computed numerically for a measurement time $t=10^2$  computed numerically for a periodically driven run-and-tumble particle with $\gamma=2$, $\Omega_0=1$, $\omega_d=1$ and for $v(t)=3(1+1/2\cos(\omega_d t)+1/3\cos(3\omega_d t))$. This even cumulant displays both a continuous background as well as spikes for $\omega_c$ corresponding to the Fourier frequencies of $v(t)^4$.}
	\label{Fig:cum_4_rtp}
\end{figure}

We next consider a simple extension of the model discussed in the previous example to characterise a particle subject to telegraphic noise confined in a time varying potential, i.e. 
\be
\dot x(t)=-\Omega(t) x(t)+v(t)\sigma(t)\;,\;\;x(0)=0\;,
\ee
where $\sigma(t)=\pm 1$ is a telegraphic noise with $\moy{\sigma(t)}=0$ and $\moy{\sigma(t_1)\sigma(t_2)}=e^{-2\gamma|t_2-t_1|}$. The noise $\sigma(t)$ is not Gaussian in this framework, which in turn renders the alternating charge $Q_{\omega_c; x}(t)$ not Gaussian either. Hence
\be
{\cal Q}_{p}^x(\omega_c)\neq 0\;,\;\;p>2\;.
\ee
In Fig. \ref{Fig:cum_4_rtp}, we have plotted the fourth order rescaled cumulant ${\cal Q}_4^{{\rm b},x}(\omega_c)$ for this specific model, showing indeed that it converges to a non-zero value.

At any time $t$, the process $(x,\sigma)$ is Markovian and this problem is therefore covered by the general theory exposed in the previous sections. The propagator $G_{\sigma,\sigma_0}(x,t|x_0,t_0)$ satisfies the equation
\begin{align}
\partial_t G_{\sigma,\sigma_0}=&-\sigma v(t)\partial_x G_{\sigma,\sigma_0}+\Omega(t)\partial_x( x G_{\sigma,\sigma_0})\label{FP_x_sig}\\
&+\gamma \left(G_{-\sigma,\sigma_0}-G_{\sigma,\sigma_0}\right)\;.\nn
\end{align}
One can again obtain the value of the position at time $t$ as
\be
x(t)=\int_0^{t}d\tau\, e^{-\int_{\tau}^{t}\Omega(\tau')d\tau'}v(\tau)\sigma(\tau)\;.
\ee

\begin{figure}[h]
	\centering
	\includegraphics[width=0.9\linewidth]{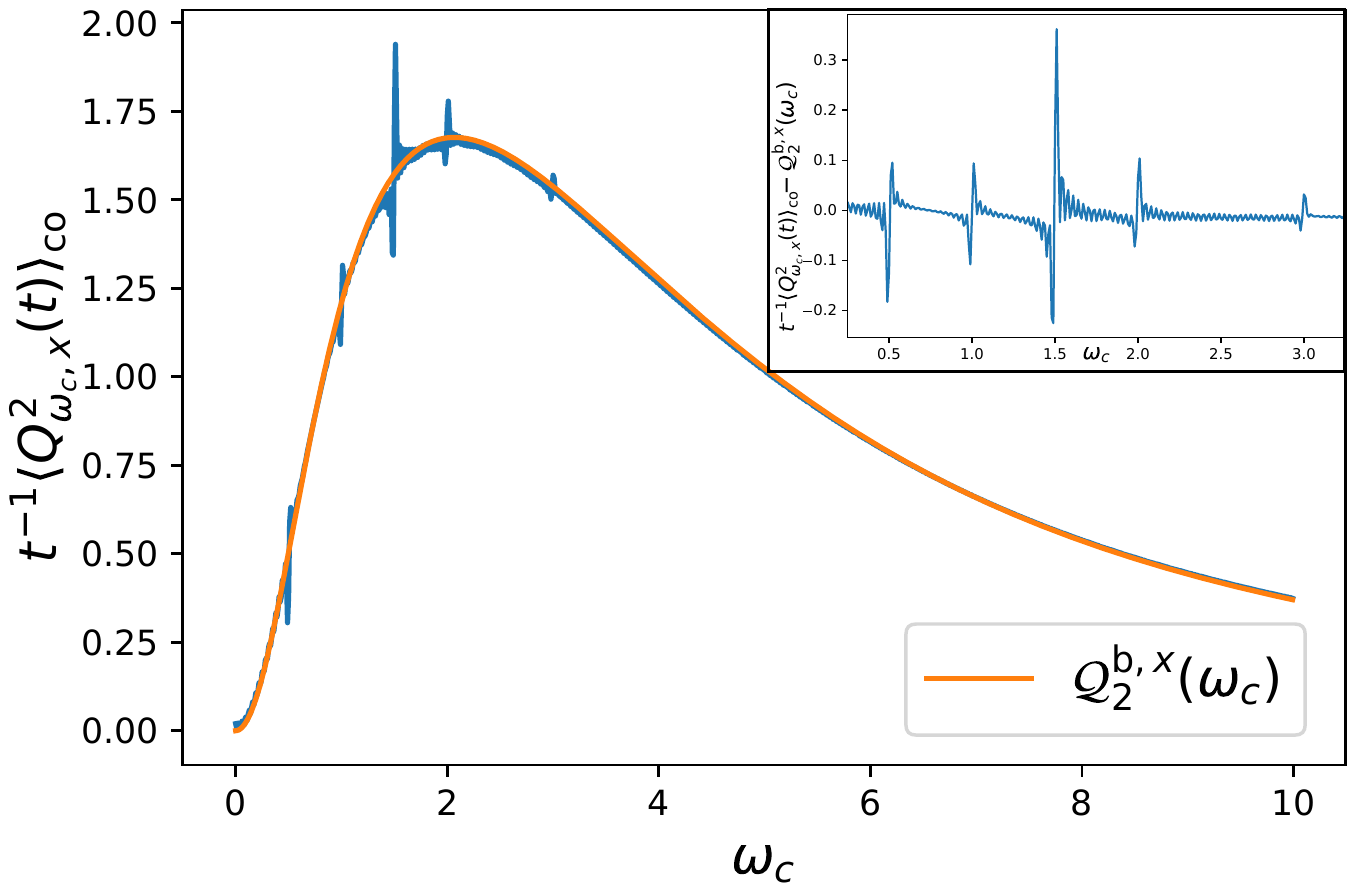}
	\caption{Comparison between the rescaled variance of the alternating charge $\moy{Q_{\omega_c,x}^2(t)}_{\rm co}/t$ (in blue) computed numerically for a measurement time $t=10^2$ for a periodically driven run-and-tumble particle with $\gamma=2$, $\Omega_0=1$, $\omega_d=1$ and for $v(t)=3(1+1/2\cos(\omega_d t)+1/3\cos(3\omega_d t))$ and the background value ${\cal Q}_2^{{\rm b},x}(\omega_c)$ (in orange) given in Eq. \eqref{var_RTP}. In inset we have plotted the difference between these two results showing that the agreement is excellent apart from the presence of spikes at values of $\omega_c$ corresponding to half the Fourier components of $v(t)^2$, i.e. for $\omega_c/\omega_d=1/2,1,3/2,2,3$. The measurement time is $t=10^2$.}
	\label{Fig:cum_2_RTP}
\end{figure}

Let us consider the special case where $\Omega(t)=\Omega_0$. We introduce the Fourier series
\be
v(t)=\sum_{k=-\infty}^{\infty} v_k\,e^{ik\omega_d t}\;.
\ee
As $\moy{x(t)}=0$, it yields immediately ${\cal Q}_1(\omega_c)=0$. The two-time connected correlation function reads instead
\begin{align}
&\moy{x(t_2)x(t_1)}_{\rm co}=e^{-\Omega_0(t_1+t_2)}\\
&\times\int_0^{t_1}d\tau_1\int_0^{t_2}d\tau_2\, e^{\Omega_0(\tau_1+\tau_2)-2\gamma|\tau_1-\tau_2|}v(\tau_1)v(\tau_2)\;. \nn  
\end{align}
Inserting the Fourier expansion of $v(t)$ result and computing explicitly the integrals, we identify this equation for large but finite time with 
\begin{align}
&\moy{x(t_2)x(t_1)}_{\rm co}\approx\\
&\sum_{l_2=1}^{\infty}e^{-\lambda_{l_2}(t_2-t_1)}\sum_{k_1,k_2=-\infty}^{\infty}C_{0,l_2}^{k_1,k_2}e^{i\omega_d(k_1 t_1+k_2 t_2)}\;.\nn
\end{align}
We thus obtain the Floquet spectrum for this model with $\lambda_0=0$ and
\be
\lambda_{1}=\min(\Omega_0,2\gamma)\leq \lambda_{2}=\max(\Omega_0,2\gamma)\;,
\ee
as well as the coefficients
\begin{align}
C_{0,2\gamma}^{k_1,k_2}=&\frac{v_{k_1}v_{k_2}}{[\Omega_0+2\gamma+ik_1\omega_d][\Omega_0-2\gamma+ik_2\omega_d]}\;,\\
C_{0,\Omega_0}^{k_1,k_2}=&\sum_{n=-\infty}^{\infty}\frac{v_{n}v_{k_1-n}\delta_{k_2,0}}{\Omega_0+2\gamma+i(k_1-n)\omega_d}\\
&\times\left(\frac{2}{2\Omega_0+ik_1\omega_d}-\frac{1}{\Omega_0-2\gamma+in\omega_d}\right)\;.\nn
\end{align}
Note that for this model, we expect a dynamical phase transition for $\Omega_0=2\gamma$ where the two Floquet eigenvalues cross. In this example, the background rescaled variance of the charge ${\cal Q}_2^{\rm b}(\omega_c)$ for $\omega_c$ different from integer and half-integer multiples of $\omega_d$ depends on the full time oscillations of $v(t)$ and reads
\begin{align}
&{\cal Q}_2^{{\rm b},x}(\omega_c)=\frac{2\gamma|v_0|^2\omega_c^2}{(\Omega_0^2+\omega_c^2)(4\gamma^2+\omega_c^2)}\label{var_RTP}\\
&+\frac{\omega_c^2}{\Omega_0^2+\omega_c^2}\sum_{n=1}^{\infty}\frac{4\gamma |v_n|^2(4\gamma^2+n^2\omega_d^2+\omega_c^2)}{(4\gamma^2+n^2\omega_d^2)^2+2(4\gamma^2-n^2\omega_d^2)\omega_c^2+\omega_c^4}\;.\nn
\end{align}
Taking the limit $\omega_c\to \infty$ for finite $\gamma$, one obtains that ${\cal Q}_2^{{\rm b},x}(\omega_c)\to 0$, which is consistent with the absence of term of the form $\partial_x^2 G$ in Eq. \eqref{FP_x_sig}. On the other hand, taking the limit $\gamma,v_n \to \infty$ with $|v_n|^2/\gamma =O(1)$, one obtains the result for the overdamped Brownian motion ${\cal Q}_2^{\rm b}(\omega_c)=D_0\omega_c^2/(\Omega_0^2+\omega_c^2)$ where $D_0=(|v_0|^2+2\sum_{n>1}|v_n|^2)/(2\gamma)$.

\subsection{Two-level system}

\begin{figure}[h]
	\centering
	\includegraphics[width=0.9\linewidth]{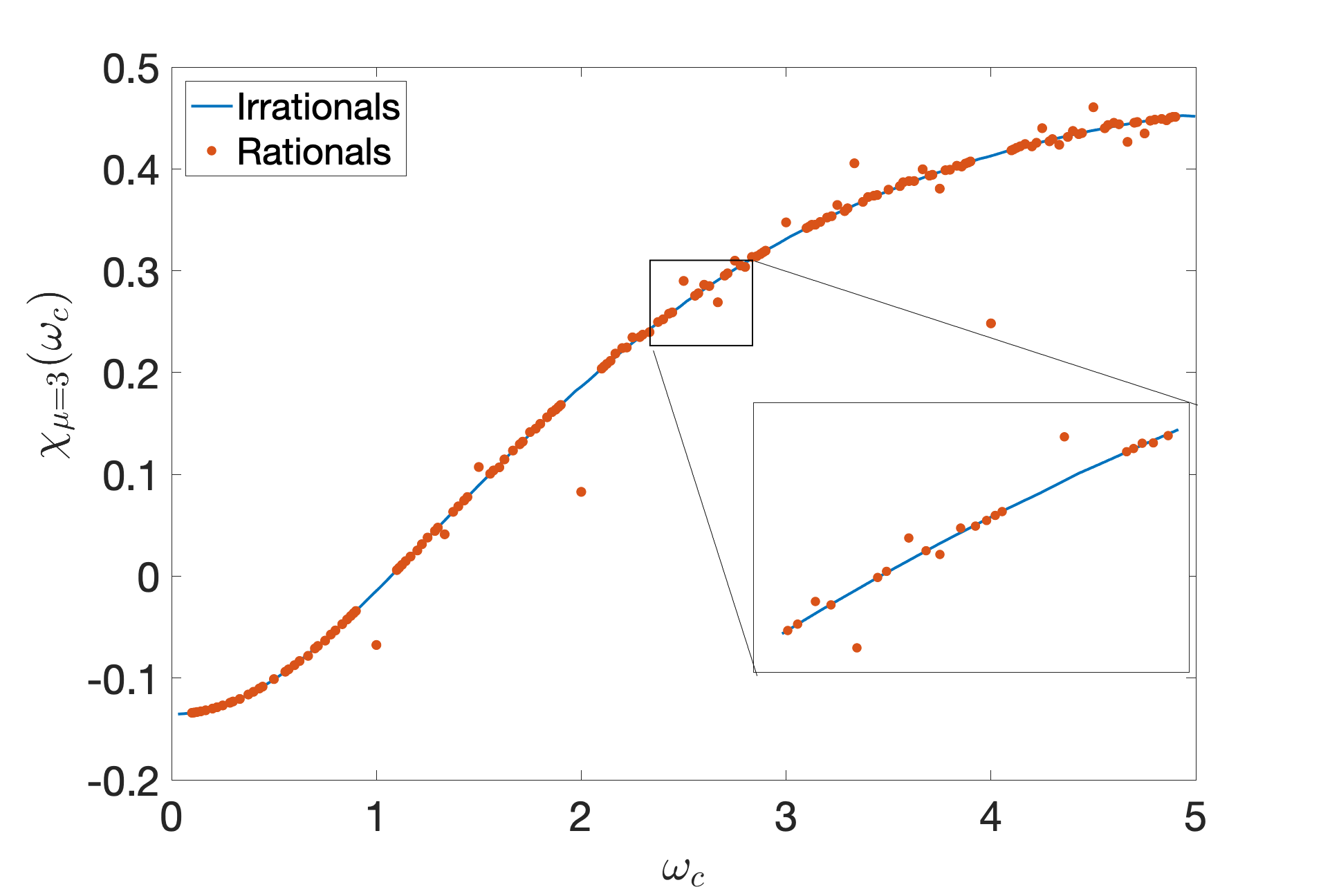}
	\caption{Plot of the CGF $\chi_{\mu}(\omega_c)$ for the 2-state system, with $\mu=3$ and $\omega_d=1$. The specific periodic driving $k_1(t)$ and $k_2(t)$ are described in equation \eqref{rates_k1_k2}. The blue line is the continuous background, calculated for incommensurate values of $\omega_c$, and the red dots where calculated at commensurate ratios. The inset shows a blowup of a small segment, where one can observe that the function $\chi_{\mu}(\omega_c)$ for rational $\omega_c$ is always away from the value of $\chi_{\mu}(\omega_c)$ at irrational $\omega_c$.}
	\label{Fig:2-state-CGF}
\end{figure}

In the previous two examples, we were able to compute the first few cumulants of the alternating charge. Achieving to obtain numerically the fourth order cumulant with enough precision was a hard task that required $\sim 10^9$ realisations. However, computing the cumulant generating function itself beyond Gaussian models, even numerically, seems a formidable task. We therefore consider next a different example, where the cumulant generating function can actually be evaluated numerically. Specifically, we consider a two level system, with a probability vector $| P(t)\rangle=(p_1(t),p_2(t))^{\rm T}$, where $p_i(t)$ is the probability for the system to be in state $i$ at time $t$. $| P(t)\rangle$ evolves in time according to a master equation,
\be
| \dot P(t)\rangle =R(t)|P(t)\rangle\;,\;\;R(t)=\begin{pmatrix}
-k_1(t)& k_2(t)\\
k_{1}(t) & -k_2(t)
\end{pmatrix}\;,\label{P_eq}
\ee
where $R(t)$ is periodic with angular frequency $\omega_d$. 
A given realisation of the process is a sequence of successive states of the system, $i_1\to i_2\to ...\to i_n$ and the corresponding sequence of times $t_k$ at which the system jumped from state $i_k$ to $i_{k+1}$. For any realisation, we define its \emph{empirical alternating charge} at frequency $\omega_c$ as 
\be
Q_{\omega_c}(t)=\sum_{k}  \Big(i_{k+1}-i_k\Big)\cos(\omega_c t_k)\;,
\ee
where the sum is over the times $t_k<t$ where the system jumps from one state to the other of the specific realization. As in the continuous state systems, we are interested in the fluctuations of this quantity as $t\to \infty$. As the system is only composed of two states, there is no cycle and therefore the direct current (DC) corresponding to $\omega_c=0$ is identically zero. Recent progress has been made on this particular system \cite{fujii2020full} and in particular an exact representation of the CGF has been obtained for the DC currents. Following the method presented in  \cite{fujii2020full} but for alternating charge (see details in Appendix \ref{app_CGF_2_level}), we obtain 
\be
\chi_{\mu}(\omega_c)=-\lim_{t\to \infty}\int_0^{t}\frac{d\tau}{t} \left[k_1(\tau)+k_2(\tau) e^{-\mu \cos(\omega_c \tau)}y_{\mu}(\tau)\right]\;.\label{chi}
\ee
In that expression, the function $y_\mu(t)$ satisfies a first order non-linear differential equation for fixed $\mu$, which reads \cite{fujii2020full}
\be
\dot y_\mu=(y_\mu-e^{\mu \cos(\omega_c t)})(k_1+k_2 e^{-\mu \cos(\omega_c t)}y_\mu)\;.\label{y_eq}
\ee
Using this particular representation, we show in Fig. (\ref{Fig:2-state-CGF}) the CGF $\chi_\mu(\omega_c)$ as a function of $\omega_c$, calculated for $\mu=3$ and the following rates:
\begin{align}
    k_1(t)&=\exp\Big(\cos(\omega_d t)-1\Big),\nonumber\\
    k_2(t)&=\exp\Big(-4\sin(5\omega_d t) + 3\sin(7\omega_dt)-2\Big)\;.\label{rates_k1_k2}
\end{align}

The blue line was calculated for the series of frequencies $\omega_c = \frac{n}{10\pi}$ for integer values of $n$, incommensurate with $\omega_d=1$. The red dots were calculated for the series of frequencies $\omega_c = \frac{n}{m}$ for  $m\in \mathbb{N}^*\leq 50$ and all $n\in \mathbb{N}\leq 6 m$, commensurate with $\omega_d$. Both the commensurate and incommensurate cases were estimated by first solving numerically the differential equation \eqref{y_eq} and then evaluating the integral in equation \eqref{chi} up to $t=2\times10^4$. 
\section{Conclusion} \label{conclu}

In this article we have computed explicitly the large time fluctuations of the alternating charge $Q_{\omega_c;\alpha}(t)$ (defined as the time integrated instantaneous alternating current in direction $\alpha$) for a  periodically driven stochastic system. We have shown that the cumulant generating function of this quantity (and as a consequence its large deviation function) is not a continuous function of the frequency $\omega_c$ of the charge. In particular, we have shown that there exists both a continuous background for any value of the charge and driving frequency, respectively $\omega_c$ and $\omega_d$, and additional peaks for commensurate pair of frequencies. This general result has been confirmed by considering some exactly solvable models. 

It would be interesting to test these fairly general results experimentally, especially on systems where current fluctuations play an important role, e.g. trapped ions setups. For this, it is important to consider the finite time version of the results.

We have focused our interest in this article on systems with a discrete gaped Floquet spectrum on a simply-connected domain, where the position of the particle are confined to a finite portion of space. It would be interesting to consider similar results for systems on a finite loop, where there might be long-lasting average currents. Some simple exactly solvable cases can be considered, hinting that the cumulants of the alternating charge behave quite differently in this case, but obtaining a general result remains a challenge at the moment.

{\it Acknowledgements :} O.R. is the incumbent of the Shlomo and Michla Tomarin career development chair, and is supported by the Abramson Family Center for Young Scientists, the Minerva and by the Israel Science Foundation, Grant No. 950/19.

\appendix

\section{Simplification of the cumulants} \label{app_q_fin_pos}

Let us first show that the CGF $\chi_{\mu}(\omega_c)$ as defined in Eq. \eqref{def_CGF} is identical to the CGF 
\be
\tilde \chi_{\mu}(\omega_c)=\lim_{t\to \infty}\frac{1}{t}\ln \moy{e^{\mu q_{\omega_c;\alpha}(t)}}\;,
\ee
where we remind that
\be
 q_{\omega_c;\alpha}(t)=\omega_c\int_0^t d\tau\, x_{\alpha}(\tau)\sin(\omega_c \tau)\;.
\ee
To show this identity, we use that
\be
Q_{\omega_c;\alpha}(t)= q_{\omega_c;\alpha}(t)+x_{\alpha}(t)\cos(\omega_c t)-x_{\alpha}(0),
\ee
where $x_{\alpha}(0)=x_0$ is fixed.
We introduce the joint PDF $P_t(q,{\bf x})$ for the respective random variables $q_{\omega_c}(t)$ and the final position ${\bf x}(t)$. We use Bayes' theorem as
\be
P_t(q,{\bf x})=P_t(q|{\bf x})P_t({\bf x})\;,
\ee
where $P_t({\bf x})=G({\bf x},t|{\bf x}_0,0)$ is the PDF of the final position ${\bf x}(t)$ and $P_t(q|{\bf x})$ is the PDF of $q_{\omega_c;\alpha}(t)$ for a fixed final position ${\bf x}(t)={\bf x}$. In the large time limit, the final position ${\bf x}(t)$ is of order $O(1)$ for the trapped systems that we consider while one naturally expects that the random variable $q_{\omega_c;\alpha}(t)=O(t)$. We thus expect the large deviation form
\be
P_t(q|{\bf x})\asymp e^{-t \varphi\left(\frac{q}{t},{\bf x}\right)}\;,
\ee
while there is no large deviation form for the final position
\be
-\lim_{t\to \infty}\frac{1}{t}\ln G({\bf x},t|{\bf x}_0,0)=\lambda_0=0\;.
\ee
We may then show that on the one hand the MGF for $Q_{\omega_c;\alpha}(t)$ reads in the large time limit
\begin{align}
&\moy{e^{\mu Q_{\omega_c;\alpha}(t)}}\\
&=e^{-\mu x_0}\int d{\bf x}\,G({\bf x},t|{\bf x}_0,0)e^{\mu x_{\alpha} \cos(\omega_c t)}\int dq\, e^{\mu q}P_t(q|{\bf x})\nn\\
&\asymp e^{-\mu x_0}\int d{\bf x}\,f_0({\bf x},t)e^{\mu x_{\alpha} \cos(\omega_c t)}\int \frac{dQ}{t}\, e^{t\left[\mu Q -\varphi(Q,{\bf x})\right]}\;,\nn
\end{align}
while on the other hand the MGF for $q_{\omega_c;\alpha}(t)$ reads
\begin{align}
&\moy{e^{\mu\, q_{\omega_c;\alpha}(t)}}\\
&=e^{-\mu x_0}\int d{\bf x}\,G({\bf x},t|{\bf x}_0,0)\int dq\, e^{\mu q}P_t(q|{\bf x})\nn\\
&\asymp e^{-\mu x_0}\int d{\bf x}\,f_0({\bf x},t)\int \frac{dQ}{t}\, e^{t\left[\mu Q -\varphi(Q,{\bf x})\right]}\;.\nn
\end{align}
Using a saddle point-approximation then naturally yields the result
\be
\chi_{\mu}(\omega_c)=\tilde\chi_{\mu}(\omega_c)=\max_{Q,{\bf x}}\left[\mu Q -\varphi(Q,{\bf x})\right]\;.
\ee
As the CGF for the two random variables $Q_{\omega_c;\alpha}(t)$ and $q_{\omega_c;\alpha}(t)$ are identical, it yields immediately the identity for the rescaled cumulants 
\be
{\cal Q}_p^{\alpha}(\omega_c)=\lim_{t\to\infty}\frac{\moy{Q_{\omega_c;\alpha}^p(t)}_{\rm co}}{t}=\lim_{t\to\infty}\frac{\moy{q_{\omega_c;\alpha}^p(t)}_{\rm co}}{t}
\ee
of arbitrary order $p$. 

\section{Expression of the third order cumulant ${\cal Q}_3^{\alpha}(\omega_c)$}\label{third}

Let us now consider the cumulant of order $p=3$. To obtain its value, we first consider the three times connected correlation function for $t_3>t_2>t_1\gg 1$
\begin{widetext}
\begin{align}
    &\moy{x_\alpha(t_1)x_\alpha(t_2)x_{\alpha}(t_3)}_{\rm co}=
    \sum_{l_2,l_3=1}^{\infty}\prod_{j=2}^3 e^{-\lambda_{l_j}(t_j-t_{j-1})}C_{0,l_2,l_3}(t_1,t_2,t_3)-\sum_{l_3=1}^{\infty}\prod_{j=2}^3 e^{-\lambda_{l_3}(t_j-t_{j-1})}C_{0,l_3}(t_1,t_3)C_{0}(t_2)\;.
\end{align}
\end{widetext}
Note that $p=3$ is the first term for which there are more than one partition of $\{1,\cdots,p\}$ that give a non-zero contribution to this connected correlation, namely $\{1,2,3\}$ and $\{1,3\}\{2\}$.
\begin{widetext}
\begin{align}
{\cal Q}_3^{\alpha}(\omega_c)=&\frac{3\omega_c^3}{4i} \sum_{k_1,k_2,k_3=-\infty}^{\infty}\sum_{\sigma_1,\sigma_2,\sigma_3=\pm 1}\sum_{l_3=1}^{\infty}\frac{\sigma_1\sigma_2\sigma_3}{\lambda_{l_3}+i(\sigma_3\omega_c-k_3\omega_d)}\delta_{\sum_{j=1}^3 k_j\omega_d,\sum_{j=1}^3 \sigma_j\omega_c}\\
&\times\left[\sum_{l_2=1}^{\infty}\frac{C_{0,l_2,l_3}^{k_1,k_2,k_3}}{\lambda_{l_2}+i\left[(\sigma_2+\sigma_3)\omega_c-(k_2+k_3)\omega_d\right]}-\frac{C_{0,l_3}^{k_1,k_3}C_{0}^{k_2}}{\lambda_{l_3}+i\left[(\sigma_2+\sigma_3)\omega_c-(k_2+k_3)\omega_d\right]}\right]\;.\nn
\end{align}
\end{widetext}
This third order cumulant is always zero if the frequencies $\omega_c$ and $\omega_d$ are incommensurate. It is only non-zero for frequencies of the form $\omega_c=n\omega_d$ and $\omega_c=n\omega_d/3$ for $n\in \mathbb{N}^*$.

\section{Expression of the fourth order cumulant ${\cal Q}_4^{\alpha}(\omega_c)$}\label{fourth}

One can use the expression of the connected four point functions for $t_4>t_3>t_2>t_1\gg 1$ with $\delta t_j=t_j-t_{j-1}$ for $j=2,3,4$, which reads
\begin{widetext}
\begin{align}
&\moy{x_\alpha(t_1)x_\alpha(t_2)x_{\alpha}(t_3)x_{\alpha}(t_4)}_{\rm co}=\\
&\sum_{l_2,l_3,l_4=1}^{\infty}\prod_{j=2}^4 e^{-\lambda_{l_j}\delta t_j}C_{0,l_2,l_3,l_4}(t_1,t_2,t_3,t_4)-\sum_{l_2,l_4=1}^{\infty}e^{-\lambda_{l_4}(\delta t_4+\delta t_3)-\lambda_{l_2}\delta t_2}C_{0,l_2,l_4}(t_1,t_2,t_4)C_0(t_3)\nn\\
&-\sum_{l_3,l_4=1}^{\infty}e^{-\lambda_{l_4}\delta t_4-\lambda_{l_3}\delta t_3-\lambda_{l_3}\delta t_2}C_{0,l_3,l_4}(t_1,t_3,t_4)C_0(t_2)\nn\\
&-\sum_{l_3,l_4=1}^{\infty}e^{-\lambda_{l_4}\delta t_4-(\lambda_{l_4}+\lambda_{l_3})\delta t_3-\lambda_{l_3}\delta t_2}\left[C_{0,l_4}(t_1,t_4)C_{0,l_3}(t_2,t_3)+C_{0,l_4}(t_2,t_4)C_{0,l_3}(t_1,t_3)\right]\nn\\
&+\sum_{l_4=1}^{\infty}\prod_{j=2}^4 e^{-\lambda_{l_4}\delta t_j}C_{0,l_4}(t_1,t_4)C_0(t_2)C_0(t_3)\;.
\end{align}
\end{widetext}
The partition of $\{1,2,3,4\}$ that contribute besides the identity are for two blocks $\{1,2,4\}\{3\}$, $\{1,3,4\}\{2\}$, $\{1,3\}\{2,4\}$, $\{2,4\}\{1,3\}$ and for three blocks $\{1,4\}\{2\}\{3\}$. Using the explicit expressions for the rates $\mu_j$'s corresponding to each partition, we obtain
\begin{widetext}
\begin{align}
&{\cal Q}_4^{\alpha}(\omega_c)=\frac{3\omega_c^4}{2} \sum_{k_1,\cdots,k_4=-\infty}^{\infty}\sum_{\sigma_1,\cdots,\sigma_4=\pm 1}\sum_{l_4=1}^{\infty}\frac{\prod_{j=1}^4\sigma_j}{\lambda_{l_4}+i(\sigma_4\omega_c-k_4\omega_d)}\delta_{\sum_{j=1}^4 k_j\omega_d,\sum_{j=1}^4 \sigma_j\omega_c}\\
&\times\left(\sum_{l_2,l_3=1}^{\infty}\frac{C_{0,l_2,l_3,l_4}^{k_1,k_2,k_3,k_4}}{\prod_{n=2}^3\left[\lambda_{l_n}+i\sum_{j=n}^4(\sigma_j\omega_c-k_j\omega_d)\right]}+\frac{C_{0,l_4}^{k_1,k_4}C_0^{k_2}C_0^{k_3}}{\prod_{n=2}^3\left[\lambda_{l_4}+i\sum_{j=n}^4(\sigma_j\omega_c-k_j\omega_d)\right]}\right.\nn\\
&-\sum_{l_3=1}^{\infty}\frac{C_{0,l_3,l_4}^{k_1,k_3,k_4}C_{0}^{k_2}}{\prod_{n=2}^3\left[\lambda_{l_3}+i\sum_{j=n}^4(\sigma_j\omega_c-k_j\omega_d)\right]}-\sum_{l_2=1}^{\infty}\frac{C_{0,l_2,l_4}^{k_1,k_2,k_4}C_{0}^{k_3}}{\left[\lambda_{l_4}+i\sum_{j=3}^4(\sigma_j\omega_c-k_j\omega_d)\right]\left[\lambda_{l_2}+i\sum_{j=2}^4(\sigma_j\omega_c-k_j\omega_d)\right]}\nn\\
&\left.-\sum_{l_3=1}^{\infty}\frac{1 }{\sum_{j=3}^4\left[\lambda_{l_j}+i(\sigma_j\omega_c-k_j\omega_d)\right]}\left[\frac{C_{0,l_4}^{k_1,k_4}C_{0,l_3}^{k_2,k_3}}{\lambda_{l_4}+i\sum_{j=2}^4(\sigma_j\omega_c-k_j\omega_d)}+\frac{C_{0,l_4}^{k_2,k_4}C_{0,l_3}^{k_1,k_3}}{\lambda_{l_3}+i\sum_{j=2}^4(\sigma_j\omega_c-k_j\omega_d)}\right]\right)\;.\nn
\end{align}
\end{widetext}

\section{Cumulant generating function for the alternating charge of the 2 level system}\label{app_CGF_2_level}

A convenient way to compute the CGF $\chi_{\mu}(\omega_c)$ is to introduce a counting vector $|P_s(t)\rangle$ (see \cite{touchette2009large} for details about this method), such that the $i^{\rm th}$ component of $|P_s(t)\rangle$ is the ensemble average of $ e^{\mu Q_{\omega_c}(t)}$ on trajectories that are in the state $i$ at time $t$. The initial condition is taken such that $|P_\mu(0)\rangle = |P(0)\rangle$ at $t=0$, and it evolves with time according to
\begin{align}
|\dot P_\mu(t)\rangle &=R_\mu(t)|P_\mu(t)\rangle\;,\label{eq_P_s_R_s}\\
R_{\mu}(t)&=\begin{pmatrix}
-k_1(t)& k_2(t) e^{-\mu \cos(\omega_c t)}\\
k_{1}(t) e^{\mu \cos(\omega_c t)} & -k_2(t)
\end{pmatrix}\;,
\end{align} 
and $R_{\mu}(t)$ is called the ``tilted rate matrix''. The vector $| P_\mu(t)\rangle$ can be (formally) expressed at any time $t$ as 
\be
| P_\mu(t)\rangle=\mathcal{T} e^{\int_0^t R_\mu(\tau)d\tau} | P_\mu(0)\rangle\label{t_o_exp}\;,
\ee
where  $\mathcal{T}$ is the time ordered product. Using this counting vector method, the cumulant generating function is given by \cite{lebowitz1999gallavotti} 
\begin{eqnarray}
\chi_{\mu}(\omega_c)&=&\ln \langle e^{\mu Q_{\omega_c}(t)}\rangle=\lim_{t\to \infty}\frac{1}{t}\ln \langle 1| P_\mu(t)\rangle\nonumber\\
 &=&\lim_{t\to \infty}\frac{1}{t}\ln \langle 1|\mathcal{T} e^{\int_0^t R_\mu(\tau)d\tau}| P(0)\rangle\;,\label{CGF_def_tilted}
\end{eqnarray}
where $\langle 1|=(1,1)$. While this analytical formula is exact, it is not very convenient as the time ordered exponential in Eq. \eqref{t_o_exp} is a complicated object and hard to compute in practice. 
To simplify the computation of the counting vector, we first introduce a dynamical invariant of the process. We then obtain an explicit expression for the counting vector $| P_\mu(t)\rangle$ in terms of this dynamical invariant and finally compute the CGF. 

The derivation below follows the derivation provided in \cite{fujii2020full}, but generalizes it for all the current's Fourier components. 

\subsection{Dynamical invariant}
A dynamical invariant corresponds to a matrix $F_\mu(t)$ which is diagonalisable, has time-independent eigenvalues that we set here to $\pm 1$ for convenience and satisfies for any time the differential equation
\be
\dot F_\mu(t)=R_\mu(t) F_\mu(t)-F_\mu(t) R_\mu(t)\;.\label{Eq_F_s}
\ee
A convenient way to parametrise a matrix satisfying the first two properties is \cite{fujii2020full}
\be
F_\mu(t)=\begin{pmatrix}
z_\mu(t)& (1+z_\mu(t))y_\mu(t)^{-1}\\
(1-z_\mu(t))y_\mu(t)& -z_\mu(t)
\end{pmatrix}\;,
\ee
with left and right time-dependent orthonormal eigenvectors $\langle r_\sigma(t)|l_\sigma'(t)\rangle=\delta_{\sigma,\sigma'}$ with $\sigma,\sigma'=\pm$ that read \cite{fujii2020full}
\begin{align}
&|l_+(t)\rangle=\begin{pmatrix}
\displaystyle 1\\
\displaystyle \frac{1-z_\mu(t)}{1+z_\mu(t)}y_\mu(t)
\end{pmatrix}\;,\label{eigenv}\\
&\langle r_{+}(t)|=\frac{1}{2}\begin{pmatrix}
\displaystyle 1+z_\mu(t)& \displaystyle \frac{1+z_\mu(t)}{y_\mu(t)}
\end{pmatrix}\;,\nn\\
&|l_-(t)\rangle=\frac{1}{2}\begin{pmatrix}
\displaystyle 1-z_\mu(t)\\
\displaystyle -(1-z_\mu(t))y_\mu(t)
\end{pmatrix}\;,\nn\\
&\langle r_{-}(t)|=\begin{pmatrix}
\displaystyle 1& \displaystyle -\frac{1+z_\mu(t)}{1-z_\mu(t)}\frac{1}{y_\mu(t)}
\end{pmatrix}\;.\nn
\end{align}
For the matrix $F_\mu(t)$ to satisfy Eq. \eqref{Eq_F_s}, the functions $y_\mu(t)$ and $z_\mu(t)$ need to satisfy the differential equations
\begin{align}
\dot y_\mu=&(y_\mu-e^{\mu \cos(\omega_c t)})(k_1+k_2 e^{-\mu \cos(\omega_c t)}y_\mu)\;,\label{y_eq}\\
\dot z_\mu=&k_2 e^{-\mu \cos(\omega_c t)}(1-z_\mu)y_\mu-k_1 e^{\mu \cos(\omega_c t)}\frac{(1+z_\mu)}{y_\mu}\;.\nn
\end{align}
Any pair of initial values $(y_\mu(0),z_\mu(0))$ gives a valid dynamical invariant. Thus, choosing such a pair defines a valid dynamical invariant. We focus here on the long-time behaviour such that the initial condition is not relevant and choose for convenience $y_\mu(0)=0$ and $z_\mu(0)=-1$. The equation for $y_\mu(t)$ has two fixed point: a stable fixed point $-k_1(t)/k_2(t)e^{\mu \cos(\omega_c t)}<0$ and an unstable fixed point $e^{\mu \cos(\omega_c t)}>0$. Note that starting from $y_\mu(0)\leq 0$, the solution $y_\mu(t)$ oscillates around the stable fixed point and remains negative at all time $t$. The equation for $z_\mu(t)$ is linear and can be solved exactly. In particular, it is easy to realise that the function $z_\mu(t)$ grows exponentially with time $t$ as $y_\mu(\tau)<0$ for any $\tau\in[0,t]$
\be
z_\mu(t)\asymp e^{-\int_0^{t}d\tau\left[y_\mu(\tau)k_2(\tau) e^{-\mu \cos(\omega_c \tau)}+\frac{k_1(\tau)}{y_\mu(\tau)} e^{\mu \cos(\omega_c \tau)}\right]}\;.
\ee
In the large time limit, the expressions of the eigenvectors in Eq. \eqref{eigenv_2} simplify
\begin{align}
&|l_+(t)\rangle\approx\begin{pmatrix}
\displaystyle 1\\
\displaystyle y_\mu(t)
\end{pmatrix}\;,\langle r_{+}(t)|\approx\frac{z_\mu(t)}{2}\begin{pmatrix}
\displaystyle 1& \displaystyle y_\mu^{-1}(t)
\end{pmatrix}\;,\nn\\
&|l_-(t)\rangle\approx\frac{z_\mu(t)}{2}\begin{pmatrix}
\displaystyle -1\\
\displaystyle y_\mu(t)
\end{pmatrix}\;,\langle r_{-}(t)|\approx\begin{pmatrix}
\displaystyle 1& \displaystyle -y_\mu^{-1}(t)
\end{pmatrix}\,.\label{eigenv_2}
\end{align}

\subsection{Expressing the counting vector and the CGF}

Coming back to the problem of finding an expression for the counting vector $|P_\mu(t)\rangle$, one can check by using Eq. \eqref{eq_P_s_R_s} together with Eq. \eqref{Eq_F_s} that the following identity holds
\be
\partial_t(F_\mu(t)| P_\mu(t)\rangle)=R_\mu(t)F_\mu(t) |P_\mu(t)\rangle\;.\label{R^s_F_s}
\ee
It is then possible to express the counting vector $|P_\mu(t)\rangle$ in the basis of the eigenvectors $|l_\pm(t)\rangle$ of $F_\mu(t)$. In this basis, one has that
\be
|P_\mu(t)\rangle=\sum_{\sigma=\pm}a_{\sigma}(t)|l_\sigma(t)\rangle\,,\,\,a_{\sigma}(t)=\langle r_{\sigma}(t)|P_\mu(t)\rangle\,.
\ee
Inserting this equation into Eqs. \eqref{eq_P_s_R_s} and \eqref{R^s_F_s}, one obtains that for $\sigma,\sigma'=\pm$
\be
(\langle r_{\sigma'}(t)|R_\mu(t)| l_\sigma(t)\rangle-\langle r_{\sigma'}(t)|\dot l_\sigma(t)\rangle)a_{\sigma}(t)=\delta_{\sigma,\sigma'}\dot a_{\sigma}(t)\,.\label{a_sig}
\ee
Inserting the expressions of the eigenvectors in Eq. \eqref{eigenv}, using the expression of the tilted rate matrix and the differential equations \eqref{y_eq}, one can then check that $\langle r_{\sigma'}(t)|R_\mu(t)| l_\sigma(t)\rangle=\langle r_{\sigma'}(t)|\dot l_\sigma(t)\rangle$ for $\sigma\neq \sigma'$. On the other hand, Eq. \eqref{a_sig} yields
\begin{align}
|P_\mu(t)\rangle=&\sum_{\sigma=\pm}a_{\sigma}(0)e^{\int_0^t d\tau\phi_{\sigma}(\tau)}|l_\sigma(t)\rangle\;,\label{P_s}\\
\phi_{\pm}(\tau)=&\langle r_{\pm}(\tau)|R_\mu(\tau)| l_\pm(\tau)\rangle-\langle r_{\pm}(\tau)|\dot l_\pm(\tau)\rangle\;.
\end{align}
We can now express $|P_\mu(t)\rangle$ in the large time limit as a function of the two functions $z_\mu(t)$ and $y_\mu(t)$. Inserting the expressions of the eigenvectors  in Eq. \eqref{eigenv_2}, using the expression of the tilted matrix $R_\mu(t)$ and the differential equations in Eq \eqref{y_eq}, one obtains at large time
\be
|P_\mu(t)\rangle \asymp e^{-\int_0^{t} d\tau\left[k_1(\tau)+k_2(\tau) e^{-\mu \cos(\omega_c \tau)}y_\mu(\tau)\right]}\;.
\ee
Finally, using Eq. \eqref{CGF_def_tilted}, we obtain the explicit expression of the CGF as given in Eq. \eqref{chi} of the main text.

\bibliography{CurrentFluc}

\end{document}